\documentclass[useAMS,usenatbib,usegraphicx]{mn2e}

\usepackage{aas_macros,url}

\title[Radar Observation of 2011 October Draconid Outburst]{Radar Observations of the 2011 October Draconid Outburst}
\author[Q.-Z.~Ye et al.]
  {Quanzhi~Ye$^1$\thanks{E-mail: qye22@uwo.ca},
  Peter~G.~Brown$^{1,2}$, Margaret~D.~Campbell-Brown$^1$, and\newauthor Robert~J.~Weryk$^1$\\
  $^{1}$Department of Physics and Astronomy, The University of Western Ontario, London, Ontario, N6A 3K7 Canada\\
  $^{2}$Centre for Planetary Science and Exploration, The University of Western Ontario, London, Ontario, N6A 5B8 Canada}
\begin{document}

\date{Accepted 1970 January 1. Received 1970 January 1; in original form 1970 January 1}

\pagerange{\pageref{firstpage}--\pageref{lastpage}} \pubyear{2013}

\maketitle

\label{firstpage}

\begin{abstract}
A strong outburst of the October Draconid meteor shower was predicted for 2011 October 8. Here we present the observations obtained by the Canadian Meteor Orbit Radar (CMOR) during the 2011 outburst. CMOR recorded 61 multi-station Draconid echoes and 179 single-station overdense Draconid echoes (covering the magnitude range of $+3\leq M_V \leq+7$) between 16-20h UT on 2011 October 8. The mean radiant for the outburst was determined to be $\alpha_g=261.9^{\circ} \pm 0.3^{\circ}$, $\delta_g=+55.3^{\circ} \pm 0.3^{\circ}$ (J2000) from observations of the underdense multi-station echoes. This radiant location agrees with model predictions \citep[e.g.][]{vau11} to $\sim1^{\circ}$. The determined geocentric velocity was found to be $\sim 10-15\%$ lower than the model value ($17.0-19.1$ $\mathrm{km \cdot s^{-1}}$ versus 20.4 $\mathrm{km \cdot s^{-1}}$), a discrepancy we attribute to undercorrection for atmospheric deceleration of low density Draconid meteoroids as well as to poor radar radiant geometry during the 
outburst peak. The
mass index at the time of the outburst was determined to be $\sim 1.75$
using the amplitude distribution of underdense echoes, in general agreement with the value of $\sim 1.72$ found using the diffusion-limited durations of overdense Draconid echoes. The relative flux
derived from overdense echo counts showed a similar variation to the meteor rate derived from visual observations. We were unable to measure the peak flux due to the high elevation of the radiant (and hence low elevation of specular Draconid echoes). Using the observed speed and electron line density measured by CMOR for all underdense Draconid echoes as a function of height as a constraint, we have applied the ablation model developed by \citet{cam04}. From these model comparisons, we find that Draconid meteoroids at radar sizes are consistent with a fixed grain number $n_{grain}=100$ and a variable grain mass $m_{grain}$ between $2\times10^{-8}~\mathrm{kg}$ and $5\times10^{-7}~\mathrm{kg}$, with bulk and grain density of $300~\mathrm{kg \cdot m^{-3}}$ and $3~000~\mathrm{kg \cdot m^{-3}}$, respectively. One particular Draconid underdense echo displayed well-defined Fresnel amplitude oscillations at four stations. The internal synchronization allowing us to measure absolute length as a function of time by
combining the
absolute timing offsets between stations. This event showed clear deceleration and modelling suggests that the number of grains for this meteoroid was of the order of 1 000 with grain masses between $10^{-10}$ and $10^{-9}$ kg, and a total mass of $2\times10^{-6}$ kg.
\end{abstract}


\section{Introduction}

\subsection{A Brief History of Draconids}

Although quiet in most years, the October Draconid meteor shower (sometimes shorten to ``Draconids'', and also referred to in older literatures as the ``Giacobinids'', after its parent body 21P/Giacobini-Zinner) has produced several of the most spectacular meteor storms in modern astronomical history. It is also among the very first (if not the first) meteor shower to be predicted on the basis of the parent comet's proximity to Earth's orbit prior to the shower being widely observed.

21P/Giacobini-Zinner was discovered by Michel Giacobini in 1900, before any modern observations of the Draconids. Because of limited observations, the exact orbit of the comet was unknown until Ernst Zinner recovered it in 1913 \citep{kro08}. The orbital period was determined to be $\sim6.5~\mathrm{yr}$, identifying the comet as a Jupiter-family comet, a major comet family that occasionally experience strong perturbations from Jupiter.

The possibility of a meteor shower generated by the comet was first proposed by \citet{dav15}, years before \citet{den27} first observed the meteor shower in either 1920 or 1926 (Denning only saw five meteors in 1920, therefore the linkage is questionable). The maximum zenith hourly rate (ZHR) in 1926 was reported to be 17 \citep{den27}. Before the historic Draconid meteor storm in 1933, no other definitive observations of the shower had been reported.

The 1933 Draconid meteor storm was largely a surprise as no specific prediction suggested an unusually intense shower that year. The meteor rate reported by European and American observers was between $2~400$ and $5~400~\mathrm{hr^{-1}}$ \citep{wat34,nij35}, while more recent estimates have suggested $10~000\pm2~000$ \citep{jen95}. The apparent radiant was measured to be near $\alpha_p=266^{\circ}$, $\delta_p=+59^{\circ}$ by \citet{nij35}, with a relatively large diameter ($\sim 10^{\circ}$). \citet{fis34} suggested as many as six possible historical sightings before 1920, with the earliest sighting dating back to Chinese observations in 585 A.D., although later studies do not support that hypothesis \citep[e.g.][]{imo58}\footnote{The Chinese observations were quoted from Sui Shu (the Book of Sui), Vol. XXI, Sect. 3, which recorded that a meteor outburst took place on Sept. 23, 585 A.D., approximately corresponding to $\lambda_{\odot}=202^{\circ}$ in J2000 epoch. Most studies
mentioning this event quote \citet{bio48}: ``...hundreds of meteors fled in every direction.'' But there was one sentence from the original Sui Shu text omitted by Biot: ``(astrologers) divined and said: `small meteors fleeing in every direction means common people is moving around.' '', apparently indicating the meteors were mostly faint. \citet{jen06} suggested a linkage to the Orionids; but given the fact that the moon was at its last quarter, and the Orionid radiant didn't rise until midnight, the event was unlikely to take place other than in the early evening as the moon would wash out most faint meteors. Since the observations were conducted in the capital of Sui, which was Chang'an (today's Xi'an); at a latitude of $34^{\circ}$ N, the radiant of the Draconids would be high enough for the described viewing geometry (``fled in every direction''), therefore a linkage of the 585 A.D. observation to Draconids cannot be ruled out.}.

A Draconid outburst at 21P/Giacobini-Zinner's 1939--1940 apparition was predicted by \citet{wat39}, but no activity was observed in either year. However, Watson made a successful outburst prediction for 21P/Giacobini-Zinner's 1946 apparition \citep{wat46}. The Draconid meteor storm again favored American and European observers, with maximum activity at around 10h UT on October 10, 1946 and a ZHR of $12~000\pm3~000$ \citep{jen95}. It was also a landmark in meteor studies, not only because it was the first time that radars were deployed to observe a meteor shower \citep{hey47,lov47,ste47}, but also because during this meteor storm many precise photographic records of bright Draconid meteors were obtained \citep[e.g.][]{jac50}.

The 1946 Draconid meteor storm was the second and the last recorded Draconid meteor storm in the 20th century. 21P/Giacobini-Zinner's apparitions in 1985 and 1998 were accompanied with sub-storm outbursts with a maximum ZHR of $\sim700$ \citep{lin87,arl98}. Moderate activity was detected by radar in 1952, 1972, and 2005 \citep{dav55,hug73,cam06}, but the associated visual activity was either unreported (1952), very weak (1972), or moderate (2005) with ZHR$\sim40$ \citep{mal72,cam05}. In other years, the hourly rates of Draconids were no more than a few meteors per hour.

\subsection{Dynamical Modelling of the Draconid Stream}

An important aspect of meteor physics is that it can reveal physical properties of small solar system bodies without the cost of a spacecraft mission. Meteoroid streams are formed from ejections occurring during multiple perihelion passages of a parent body as well as occasionally through catastrophic break-ups. The meteoroids released during each ejection epoch will undergo a slightly different orbital evolution compared to the parent body, depending on their mass and ejection state, but generally they are diluted into the background (annual) stream which is usually too sparse to allow a statistically sufficient sampling of shower meteoroids and typically includes meteoroids from many ejection eras. However, recently ejected material from the parent body may encounter the Earth and produce a meteor outburst, often comprising material from a single ejection epoch. Observations of such outbursts allow us to directly sample a large number of meteoroids of known age from the parent body. The interaction of these
meteoroids with the Earth's atmosphere can reveal clues about their properties, such as mass and chemical composition, which are directly related to the parent body. For example, the mismatch of visual and radar activity in the 2005 Draconid outbursts may suggest that the 2005 return was dominated by an abundance of sub-visual meteoroids from the 1946 perihelion passage of 21P/Giacobini-Zinner \citep{cam05}.

The dust trail theory of meteor outbursts/storms developed in the 1990s turned out to be a robust way to understand the linkage between outbursts of meteor showers and both the activity and ejection age from the parent body \citep{eme92,ash99,arl99}. However, to apply such a theory with confidence, sufficient observations of the meteor shower itself and its parent body are needed as constraints. Virtually, no meteor showers can satisfy this requirement except the Leonids, mostly because of a lack of significant trail encounters and outbursts that might allow us to accumulate sufficient observations about individual dust trails.

Prior to the late 1990's, potential Draconid outbursts were predicted by considering the distance between the orbits of 21P/Giacobini-Zinner and the Earth, the time difference between the arrival of the two bodies to the intersection point, and the geometry of comet's perihelion relative to the Earth's orbit. The predictions of Draconid outbursts in 1946, 1985 and 1998 using this method were in good agreement with observations \citep{wat46,spa85,lan97}, while the predictions for 1952, 1959, 1972 and 1979 did not agree. This reflects the value of applying the dust trail model to young, outburst streams such as the Draconids.

\subsection{Outburst Predictions for 2011}

\citet{sat03} first applied the dust trail theory to the Draconids and studied the trail encounter situations in 1998 and 1999, while \citet{cam06} adopted a stream model proposed by \citet{vau05} to study the unexpected outburst of the Draconids in 2005. Further numerical simulations were carried out by a number of researchers (see Table~\ref{tbl0} for a summary), and Earth's encounter of 21P/Giacobini-Zinner's ejections from 1866 to 1907 (hereafter the 1866--1907 ``trails'') were predicted to occur between 16--20h UT, October 8, 2011. The associated peak visual rate predictions ranged from $50~\mathrm{hr^{-1}}$ \citep{mas11} to $600~\mathrm{hr^{-1}}$ \citep{vau11}. Almost all observing techniques, including airborne missions, were deployed for the outburst, and preliminary analyses indicate that the outburst arrived more or less as predicted \citep{vau11b}. At the Canadian Meteor Orbit Radar (CMOR), for example, a meteor rate of $\sim 140~\mathrm{hr^{-1}}$ was observed around 19h UT, October 8, 2011, 
indicated an unusually high activity comparing to past non-outburst years ($\sim 26~\mathrm{hr^{-1}}$ and $\sim 39~\mathrm{hr^{-1}}$ in 2009 and 2010, for example).

\begin{table*}
 \centering
  \caption{Summary of Draconid 2011 outburst predictions of time and radiant position for individual trail encounters. The forecast data from Lyytinen and Moser are quoted from \texttt{http://draconids.seti.org/} (accessed Mar. 17, 2013).}
  \begin{tabular}{cccc}
  \hline
   Trail & Reference & Predicted $\alpha_g$, $\delta_g$ & Predicted maximum \\
         &            & (J2000) & (Oct. 8, UT) \\
 \hline
1866 & \citet{vau11} & $263.3^{\circ}$, $+55.3^{\circ}$ & 16:13 \\
\hline
1873 & \citet{vau11} & $263.2^{\circ}$, $+55.4^{\circ}$ & 16:29 \\
\hline
1880 & \citet{vau11} & $263.2^{\circ}$, $+55.4^{\circ}$ & 16:53 \\
& \citet{sat03} &  & 19:04 \\
\hline
1887 & \citet{vau11} & $263.2^{\circ}$, $+55.4^{\circ}$ & 17:25 \\
& \citet{sat03} &  & 17:05 \\
& Lyytinen &  & 17:02 \\
& \citet{mas11} & $263.2^{\circ}$, $+55.4^{\circ}$ & 17:04 \\
\hline
1894 & \citet{vau11} & $263.2^{\circ}$, $+55.4^{\circ}$ & 18:45 \\
& \citet{mas11} & $263.3^{\circ}$, $+55.5^{\circ}$ & 18:06 \\
\hline
1900 & \citet{vau11} & $263.2^{\circ}$, $+55.8^{\circ}$ & 20:01 \\
& \citet{sat03} &  & 20:36 \\
& Moser &  & 19:52 \\
& Lyytinen &  & 20:12 \\
& \citet{mas11} & $263.3^{\circ}$, $+55.8^{\circ}$ & 20:42 \\
\hline
1907 & \citet{vau11} & $262.5^{\circ}$, $+55.4^{\circ}$ & 19:26 \\
& \citet{sat03} &  & 19:59 \\
& \citet{mas11} & $263.2^{\circ}$, $+55.8^{\circ}$ & 21:05 \\
\hline
\end{tabular}
\label{tbl0}
\end{table*}

In this paper, we present the observation and analysis of the 2011 Draconid outburst as recorded by the CMOR. We focus on the following topics:

\begin{enumerate}
  \item characteristics of the outburst, such as radiant, velocity distribution, mass distribution, and flux variation;
  \item comparison with dynamical stream model predictions;
  \item meteoroid structure as revealed through ablation modeling.
\end{enumerate}

\section{Radar Instrumentation}

\subsection{The CMOR System}

\begin{figure}
\includegraphics[scale=.3]{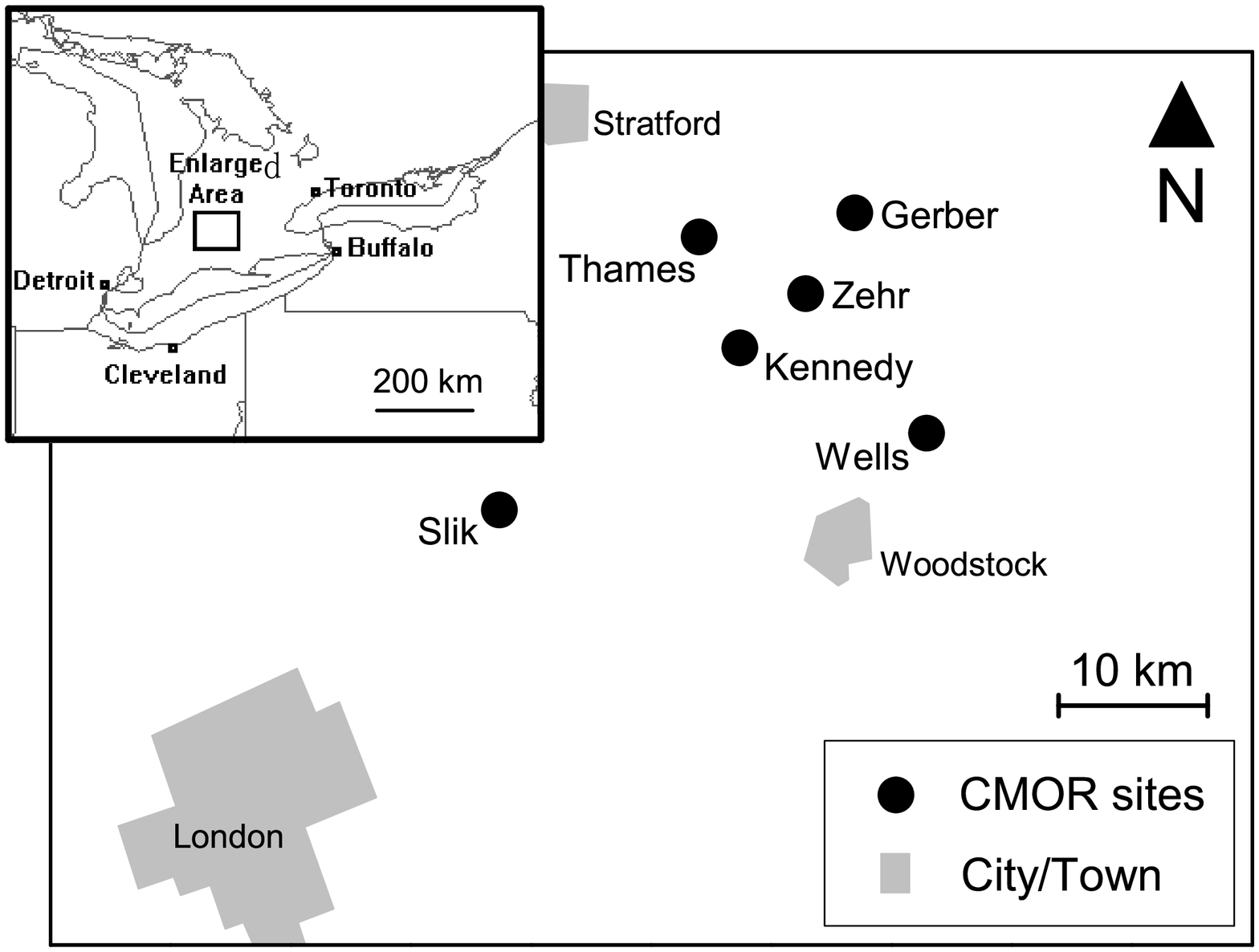}
\caption{Location and geographic distribution of the main CMOR station (Zehr) and other remote sites as of October 2011.}
\label{fig-layout}
\end{figure}

The Canadian Meteor Orbit Radar (CMOR) is an interferometric backscatter meteor radar located near London, Ontario, Canada that is designed to observe meteor echoes and perform basic analysis continuously and automatically. It is based on the commercially available SKiYMET system \citep[e.g.][]{hoc01} with some modifications to optimize for astronomical meteor echo detection \citep[e.g.][]{jon05,wer12}. Currently it consists of six sites (Figure~\ref{fig-layout}) and operates at 29.85 MHz at 12 kW peak power (Table~\ref{tbl-cmorfact})\footnote{The system also operates in 17.45 and 38.15 MHz, but only data at 29.85 MHz is used in this study.}. The radar detects meteors through reflection of a transmitted pulse from the ionization trail left behind during meteor ablation and subsequently received after specular reflection by receivers. Meteors observed by the radar from the main station (Zehr) are always $90^{\circ}$ from the apparent radiant. Because of this geometry, the underdense echo rate for radar observations has a secondary 
minimum when
the radiant has a zenith angle $<\sim20^{\circ}$ due to lower elevations and larger ranges to the echo, as opposed to visual, photographic and video meteor observation, where the minimum in apparent rates typically occurs when the radiant is near the horizon.

\begin{table*}
 \centering
  \caption{Basic specification of the 29.85 MHz CMOR system, adapted from \citet{wer12}.}
  \begin{tabular}{cc}
  \hline
  Parameter & Value \\
  \hline
   Frequency & 29.85 MHz \\
   Range interval & 15--255 km \\
   Range resolution & 3 km \\
   Pulse frequency & 532 Hz \\
   Peak power & 12 kW \\
   Noise floor & -107 dBm \\
   Dynamic range & 33 dB \\
   Beam size & $55^{\circ}$ at -3 dB point \\
\hline
\end{tabular}
\label{tbl-cmorfact}
\end{table*}

\begin{figure}
\includegraphics[scale=.3]{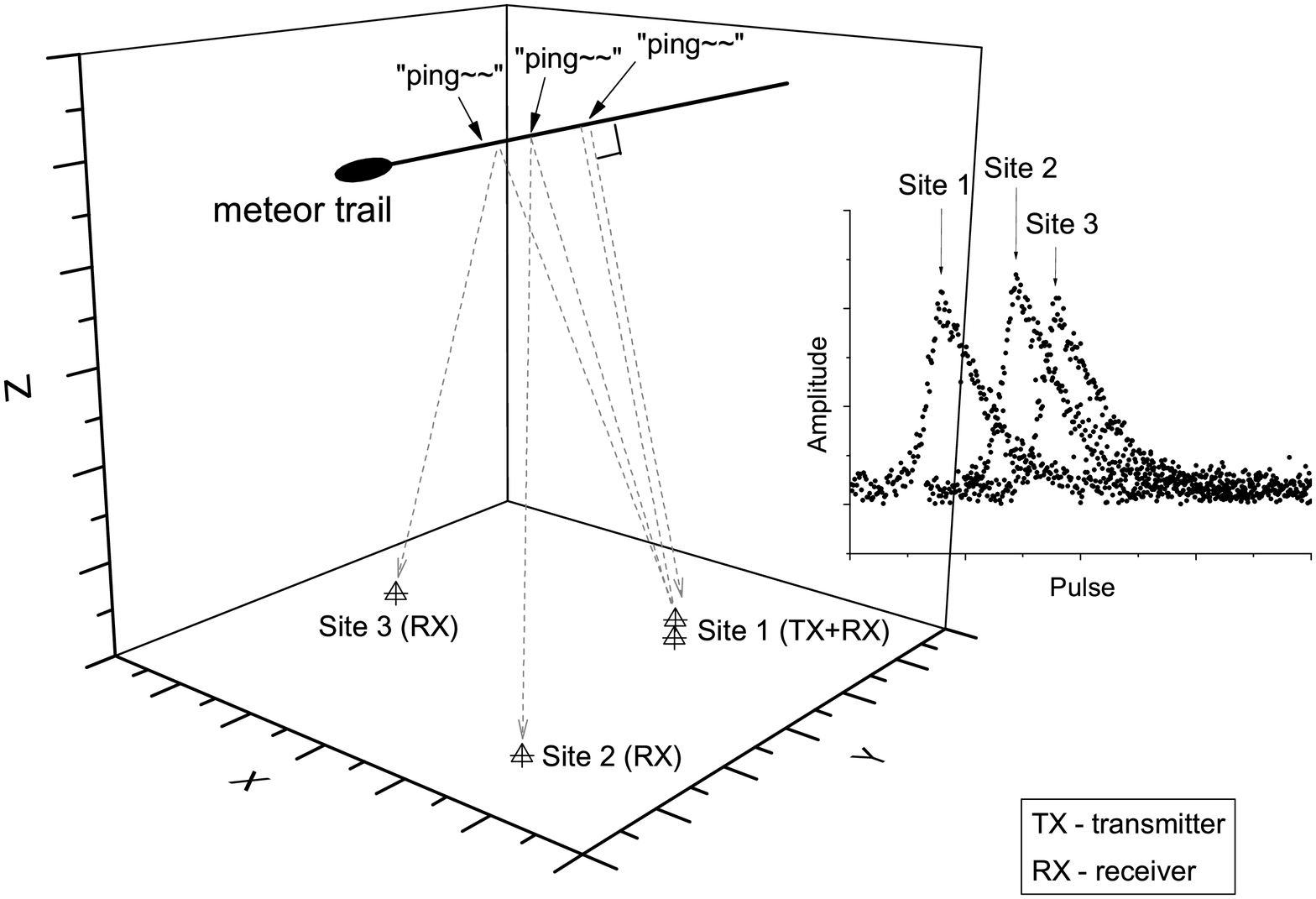}
\caption{Simplified example of how CMOR measures meteor trajectories. In this example, three radar sites detect signals reflected from the meteor trail at different points as the meteoroid moves in the atmosphere. The time differences between the three observations, together with the interferometric direction measured from the main site, can be used to construct the trajectory of the meteor.}
\label{how}
\end{figure}

If the echo from a meteor is detected at $N$ sites ($N\geq3$), we can record $N$ specular scattering positions along the meteor trail, allowing us to measure the trajectory of the meteor using a time-of-flight (\texttt{tof}) algorithm \citep[Figure~\ref{how}; see also][]{jon05,wer12}. However, the uncertainty in such a trajectory largely depends on echo strength and geometry. If the echo is weak, determination of the time of occurrence (or ``time pick'') will be difficult. Since the specular locations along a trail may coincide from different sites depending on the geometry of the meteor trajectory, the time difference between these sites may be small, and small uncertainties of the time pick may result in significant errors to the trajectory.

Meteor echoes detected by CMOR are automatically processed to remove bad detections and to correlate common events across the sites. The selection algorithms are derived from manual examination of thousands of echoes. Generally, echoes that last $<\sim 4$ s, feature a clear, sharp rise and a gradual decline, and a single maximum will be considered as a ``good'' echo, while bumpy and/or noisy echoes, such as the lower one shown in Figure~\ref{fig-echo-sample}, will be considered as a ``bad'' echo and may be rejected by the algorithm. Details of this process can be found in \citet{jon05,bro08,bro10} and \citet{wer12}.

\subsection{Echo Types}

CMOR is composed by a series of Yagi 2-element receivers and 3-element transmitters \citep{bro08}, allowing the radar to detect meteors appearing almost anywhere in the sky. The gain distribution resembles a bubble rather than a uniform distribution. This not only affects the number of meteors detected with respect to different ranges, but also affects the determination of the physical properties of individual meteoroids. One of the most significant properties of any specular radar echoes is the trail type of the meteor, namely underdense or overdense.

An underdense trail occurs when radio waves scatter from all the individual the electrons in the trail. In contrast, an overdense echo occurs when the radio wave cannot completely penetrate the meteor trail due to the trail plasma frequency being higher than the radar wave frequency. Strictly speaking, the boundary between underdense and overdense echoes is a continuum; in between is transition echoes, which can exhibit characteristics from the other two types, but for simplicity we also consider them as overdense echoes in this study.

Visually, the amplitude-time series of an overdense echo will appeared as ``flat'' (i.e. does not decay) for some time until ambipolar diffusion makes the trail underdense. Since overdense echoes have higher electron line density than underdense echoes, for a fixed velocity, overdense echoes tend to be generated by a larger meteoroid. They represent a higher fraction of echoes in regions where the radar gain is low, since fainter echoes will not be observed in these regions.

Examples of underdense and overdense echoes are shown in Figure~\ref{fig-echo-sample}. Detailed theory of these two echo types is beyond the scope of this paper, but interested readers may refer to \citet[][\S8]{mck61} or \citet[][\S4]{cep98} for details. The CMOR automatic detection algorithms are tuned to accept underdense echoes but generally suppress overdense-type echoes.

\section{Observational Methodology}

\subsection{Selection of Draconid Echoes}

For our study, we first separated the Draconid meteor echoes from other echoes. To include as many Draconid meteors as possible, we use two methods to select underdense and overdense Draconid echoes respectively; this is briefly described below and summarized in Table~\ref{tbl-dataset}.

For multi-station echoes, we developed user-interactive software to view, select, and analyze the geometry and trajectory of meteors automatically detected in the raw data stream. These are mainly underdense echoes with some overdense echoes included as well.  We consider any meteor radiant within $5^{\circ}$ and $20\%$ of the predicted 1900-trail geocentric radiant, (i.e. $\alpha_g=263.2^{\circ}$, $\delta_g=+55.8^{\circ}$, J2000), and geocentric velocity $v_g=20.4~\mathrm{km \cdot s^{-1}}$ \citep{jen06}, to be Draconid candidates. The velocity constraint is broad as Draconids are more fragile than other meteor streams \citep{jac50,fio64}, leading to larger uncertainty in measured velocity as a result of deceleration with height of detection. Using this method, we selected 61 Draconid candidates, including 39 underdense echoes and 22 overdense echoes. We refer to these as the \verb"complete" dataset.

\begin{figure}
\includegraphics[scale=.5]{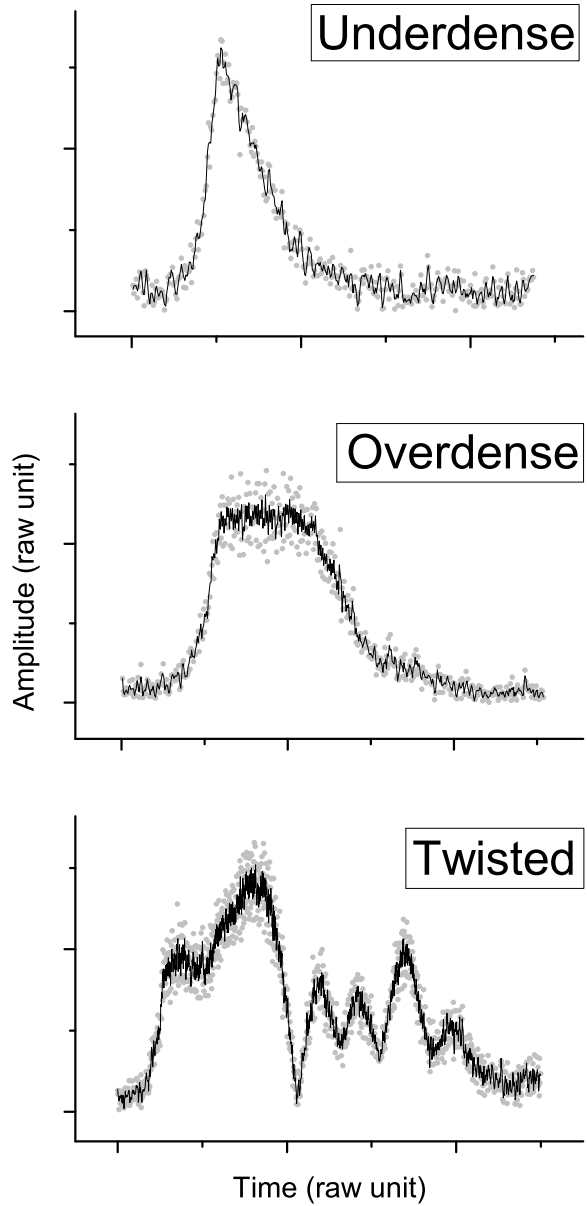}
\caption{A typical Draconid underdense echo (above), overdense echo (middle) and wind twisted overdense echo (below). We define these by the shape of their amplitude--time series (see \S2.2).}
\label{fig-echo-sample}
\end{figure}

\begin{figure}
\includegraphics[scale=.4]{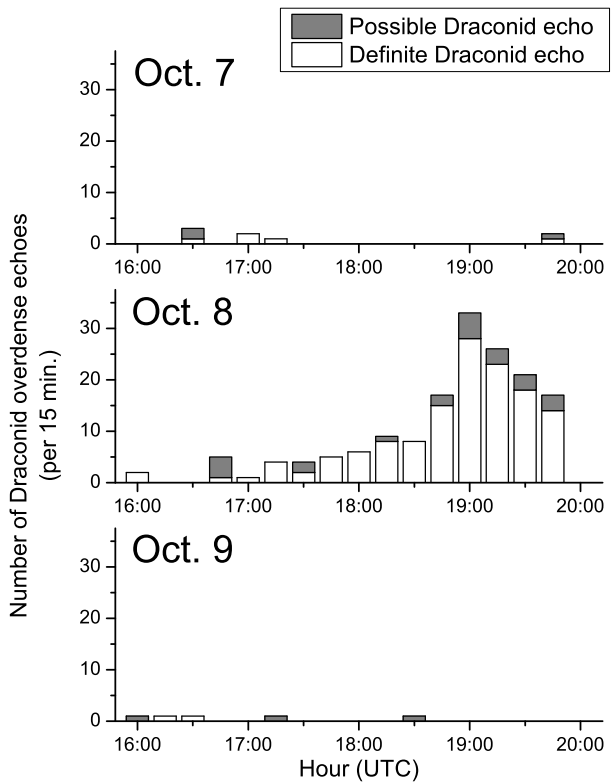}
\caption{Number-time series of definite and possible Draconid overdense echoes.}
\label{fig-od-number}
\end{figure}

Since the echo line where all Draconid meteors detectable by CMOR is at a low elevation ($<\sim 20^{\circ}$) and a larger range around the outburst ($\sim20$h UT), only the meteors which scatter the most power will be detectable (i.e. overdense echoes). However, as some non-specular overdense echoes are detected through wind-twisting or development of plasma irregularities (Figure~\ref{fig-echo-sample}), they can be difficult to detect and measure in an automated fashion. Therefore, we manually examined the raw data from the main site (Zehr) between Oct. 7--9, 2011 at 15--20h UT each day for overdense Draconid echoes. The trajectories of the trails were manually determined for cases when they were well-observed at other remote sites. This permitted us to isolate overdense Draconids. Trails with observations from less than three stations but with interferometric position consistent with Draconid meteor (i.e. at right angles to the Draconid radiant), were marked as ``possible'' Draconids. Following this method,
 we identified 148 
overdense specular
Draconid echoes as well as 31 possible Draconid echoes, which formed another sample (Figure~\ref{fig-od-number}; referred to as the \verb"overdense" dataset). This includes the 22 overdense echoes in the \verb"complete" dataset. For the relative flux analysis, the number of possible overdense echoes on October 8, 2011 is subtracted from echoes from the same periods on Oct. 7 and 9, 2011, to correct for possible sporadic overdense contamination.

\begin{table*}
 \centering
  \caption{Summary of the two datasets used in our study.}
  \begin{tabular}{ccc}
  \hline
  $N$ & \texttt{complete} dataset & \texttt{overdense} dataset \\
  \hline
   Underdense & 39 & - \\
   Overdense (definite) & 22 & 148 \\
   Overdense (possible) & - & 31 \\
  \hline
   Total & 61 & 179 \\
\hline
\end{tabular}
\label{tbl-dataset}
\end{table*}

\subsection{Uncertainties}

As the \verb"complete" dataset was measured in an automatic fashion, it is possible to objectively estimate the uncertainty in the data. This involves an estimate of the uncertainty of the position of t$_0$ (the specular point at each station) and the interferometric direction, both of which are directly measured, to estimate the error in the \texttt{tof} velocity. To do this, we used an echo simulator, which constructs a synthetic echo based on the observing geometry and strength of the measured signal assuming a Gaussian distribution of noise \citep[see][]{wer12}. Each pseudo-echo is then analyzed using the same detection/measurement algorithm that is used for real observations. A comparison is then made between the measured model results and the ``true'' (input) state to determine the expected random observational error in radiant and speed measurements, under the assumption of a perfect underdense specular echo. We ran the simulation 65 536 times\footnote{This number is arbitrary chosen so that it is 
large enough to be statistically meaningful for the Monte Carlo uncertainty analysis.} for each echo and recorded the standard deviation to be
the uncertainty in speed and radiant. We were able to use this technique on 41 echoes in the \verb"complete" dataset; for the other 20 echoes in the \verb"complete" dataset, we found it necessary to manually revise the time pick for reliable results, so no uncertainty can be estimated for these echoes. We believe the error estimation for the 41 echoes is representative of the uncertainty of the entire complete dataset.

\begin{figure}
\includegraphics[scale=.4]{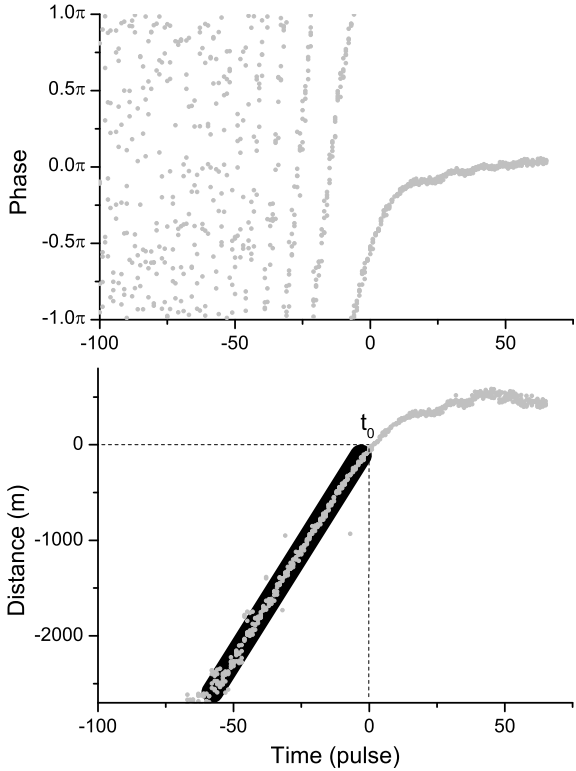}
\caption{An example of the pre-t$_0$ phase-change exhibited by a Draconid meteor echo (upper). By making use of the change of distance along the meteor trajectory, we remove the $2\pi$ jumps and make use the phase-distance relation ($d \propto \sqrt{\phi}$). The speed is found by a linear least-square fit to the meteor trail distance vs. time prior to the specular (t$_0$) point (lower). In this example, the pre-t$_0$ velocity is determined to be $21.8~\mathrm{km \cdot s^{-1}}$.}
\label{fig-pret0}
\end{figure}

\begin{figure}
\includegraphics[scale=.3]{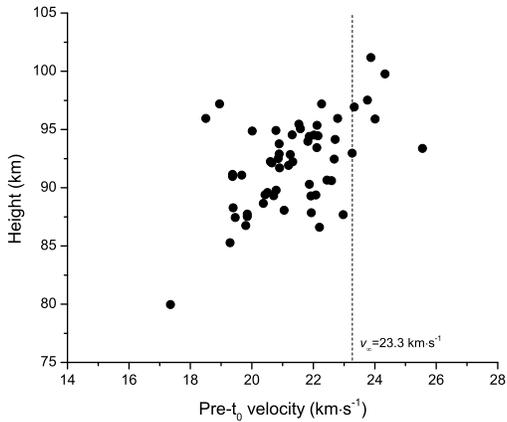}
\caption{Height distribution of meteor echoes with pre-t$_0$ determined velocities. The average deceleration trend is clearly represented by the change in speed with scattering height. The dashed vertical line represents the expected out of atmosphere speed from \citet{jen06}.}
\label{fig-pret0-ht}
\end{figure}

\begin{figure}
\includegraphics[scale=.3]{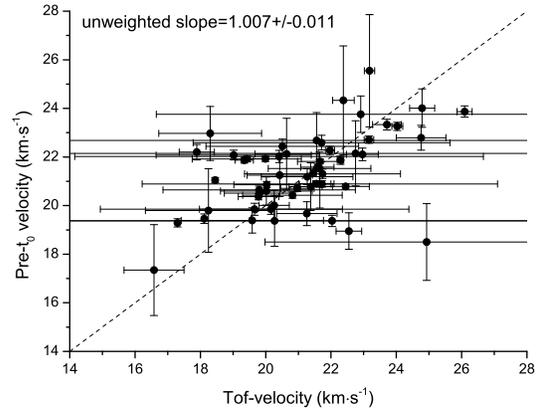}
\caption{Comparison between \texttt{tof}-derived velocities and pre-t$_0$-derived velocities. Dashed line is the 1:1 slope.}
\label{fig-pret0-tof}
\end{figure}

The velocity of each echo in the \verb"complete" dataset is determined using the \texttt{tof} method \citep{bag94}. To verify that the \texttt{tof} method yields accurate velocity measurements, we also use the Fresnel phase-time method \citep{cer97} to measure the velocities of echoes in the \verb"complete" dataset that exhibit such features. This method works as shown in Figure~\ref{fig-pret0}: when a specular meteor echo exhibits characteristic phase changes before the specular (t$_0$) point, we can determine the speed of the meteor by fitting the slope of the phase change before the t$_0$ point. Because the pre-t$_0$ measurement precision does not depend on trail geometry with respect to radar sites (as opposed to the \texttt{tof} method), it can be an independent check for the latter. We were able to manually measure the pre-t$_0$ features of 57 echoes in the \verb"complete" dataset covering a wide range of heights(Figure~\ref{fig-pret0-ht}), and found they roughly agreed with the result from the \texttt{
tof}-method where errors are largely due to geometry of \texttt{tof} and height averaging of the assumed constant speed (Figure~\ref{fig-pret0-tof}). We should note that the result from the pre-t$_0$ method is only used to broadly verify the \texttt{tof}-method, as the pre-t$_0$ Fresnel method\footnote{There is another Fresnel method called the amplitude-time method \citep[see][\S4.6.1]{cep98}; however only 7 echoes in our sample exhibited such a feature, and therefore we did not use this method. We note that the small number of Draconid echoes displaying Fresnel amplitude variations is another indication of the widespread fragmentation behavior of the Draconids.} requires subjective determination of the starting point of phase changes, so the uncertainty estimations are also subjective.

\section{Analyses, Results and Discussion}

\subsection{Radiant and Orbits}

\begin{figure*}
\includegraphics[scale=.6]{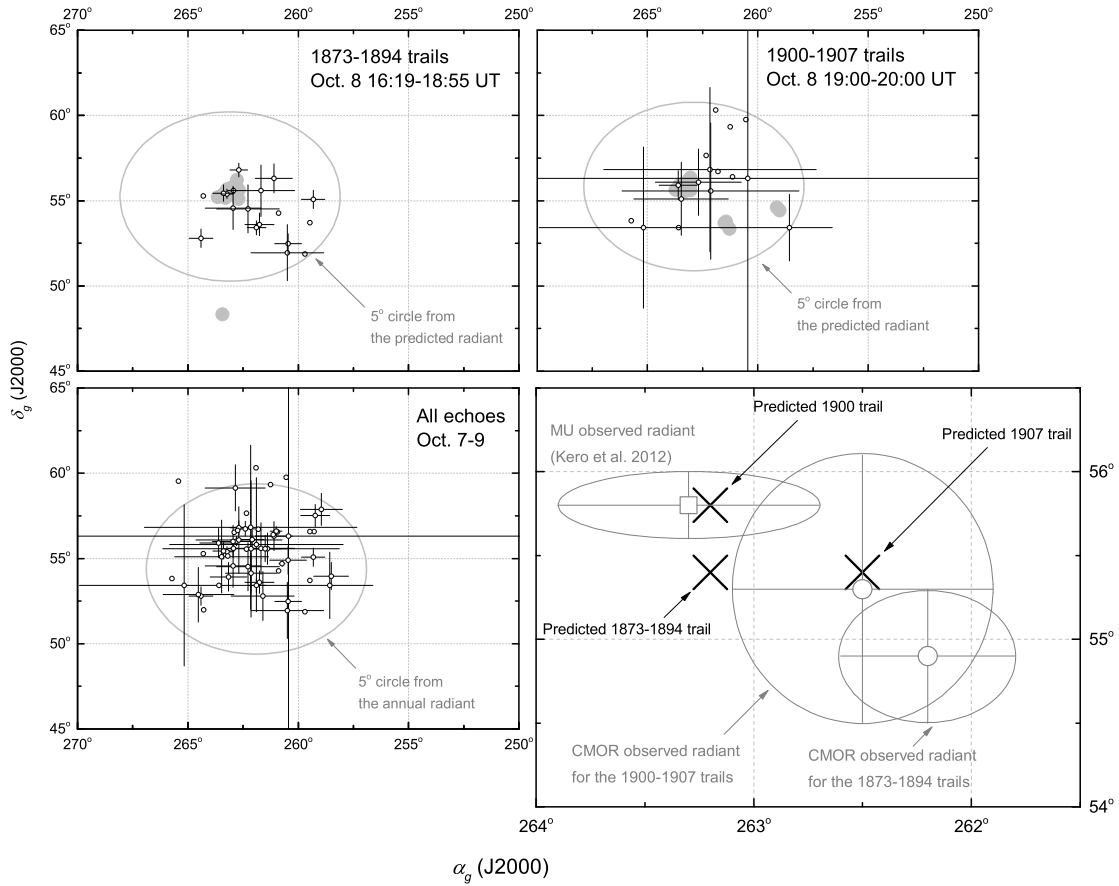}
\caption{CMOR-observed weighted mean geocentric radiant for the 1873--1894 trails (upper left), the 1900--1907 trails (upper right) and all echoes in the \texttt{complete} dataset (lower left). Uncertainty bars based on echosim results for individual radiants are plotted where applicable. Grey circles indicate a $5^{\circ}$ circle within the predicted mean radiant of the model predicted respective trail or annual radiant. Shaded grey dots in the upper two figures represent the radiant of pseudo particles in the mass range of $10^{-10}$ to $10^{-1}~\mathrm{kg}$ as simulated by \citet{vau11} (assuming a bulk density of $300~\mathrm{kg \cdot m^{-3}}$). The overall comparison between observations of CMOR and MU radar \citep{ker12}, as well as the prediction by \citet{vau11}, is given in lower right of the figure.}
\label{fig-rad-merge}
\end{figure*}

The weighted mean geocentric radiant for all echoes in the \verb"complete" dataset was determined to be $\alpha_g=261.9^{\circ}\pm0.3^{\circ}$, $\delta_g=+55.3^{\circ}\pm0.3^{\circ}$ (J2000), as shown in Table~\ref{tbl1} and Figure~\ref{fig-rad-merge}. The radiant errors reflect the scatter in individual radiants weighted by individual radiant errors found through simulation as described in \S3.2. We also examined the meteors detected around the time of two predicted main peaks (see Table~\ref{tbl0}), namely the expected peaks from the 1873--1894 and the 1900--1907 trails. We examined subsets of the \verb"complete" dataset that spanned these predicted trail passage times and found the weighted (inversely by error) geocentric radiants. Comparing these to the predictions summarized in Table~\ref{tbl0}, the radiants determined from our observations match the model prediction reasonably well ($\sim 1^{\circ}$), and are comparable to the observations of Shigaraki middle and upper atmosphere (MU) radar \citep{
ker12} (Figure~\ref{fig-rad-merge}; we note that the MU results span a large period of
time and thus represent a mixture of meteoroids from both 1873--1894 and 1900--1907 trails).

\begin{table*}
 \centering
  \caption{CMOR-observed weighted mean geocentric radiant and velocity for all detected Draconid meteors. Also shown are average radiants and speeds expected to be associated with the 1873--1894 and 1900--1907 trails based on the echo time of appearance in comparison to the model in the \texttt{complete} dataset.}
  \begin{tabular}{cccc}
  \hline
   Time period & $\alpha_g$, $\delta_g$ & $v_g$ & $N$ \\
    & (J2000) & ($\mathrm{km \cdot s^{-1}}$) & \\
 \hline
 Annual radiant by \citet{jen06} & $264.1^{\circ}$, $+57.6^{\circ}$ & 20.4 & \\
CMOR observation: Oct. 7--9 & $261.9^{\circ}\pm0.3^{\circ}$, $+55.3^{\circ}\pm0.3^{\circ}$ & $19.1\pm0.3$ & 61 \\
\hline
Vaubaillon's prediction for 1873--1894 trails & $263.2^{\circ}$, $+55.4^{\circ}$ & & \\
CMOR observation: Oct. 8 16:19--18:55UT (1873--1894 trails) & $262.2^{\circ}\pm0.4^{\circ}$, $+54.9^{\circ}\pm0.4^{\circ}$ & $18.3\pm0.4$ & 32 \\
\hline
Vaubaillon's prediction for 1900-07 trails & $262.8^{\circ}$, $+55.6^{\circ}$ & & \\
CMOR observation: Oct. 8 19:00--20:00UT (1900--1907 trails) & $262.5^{\circ}\pm0.6^{\circ}$, $+55.3^{\circ}\pm0.8^{\circ}$ & $17.0\pm0.5$ & 16 \\
\hline
\end{tabular}
\label{tbl1}
\end{table*}

\begin{figure}
\includegraphics[scale=.3]{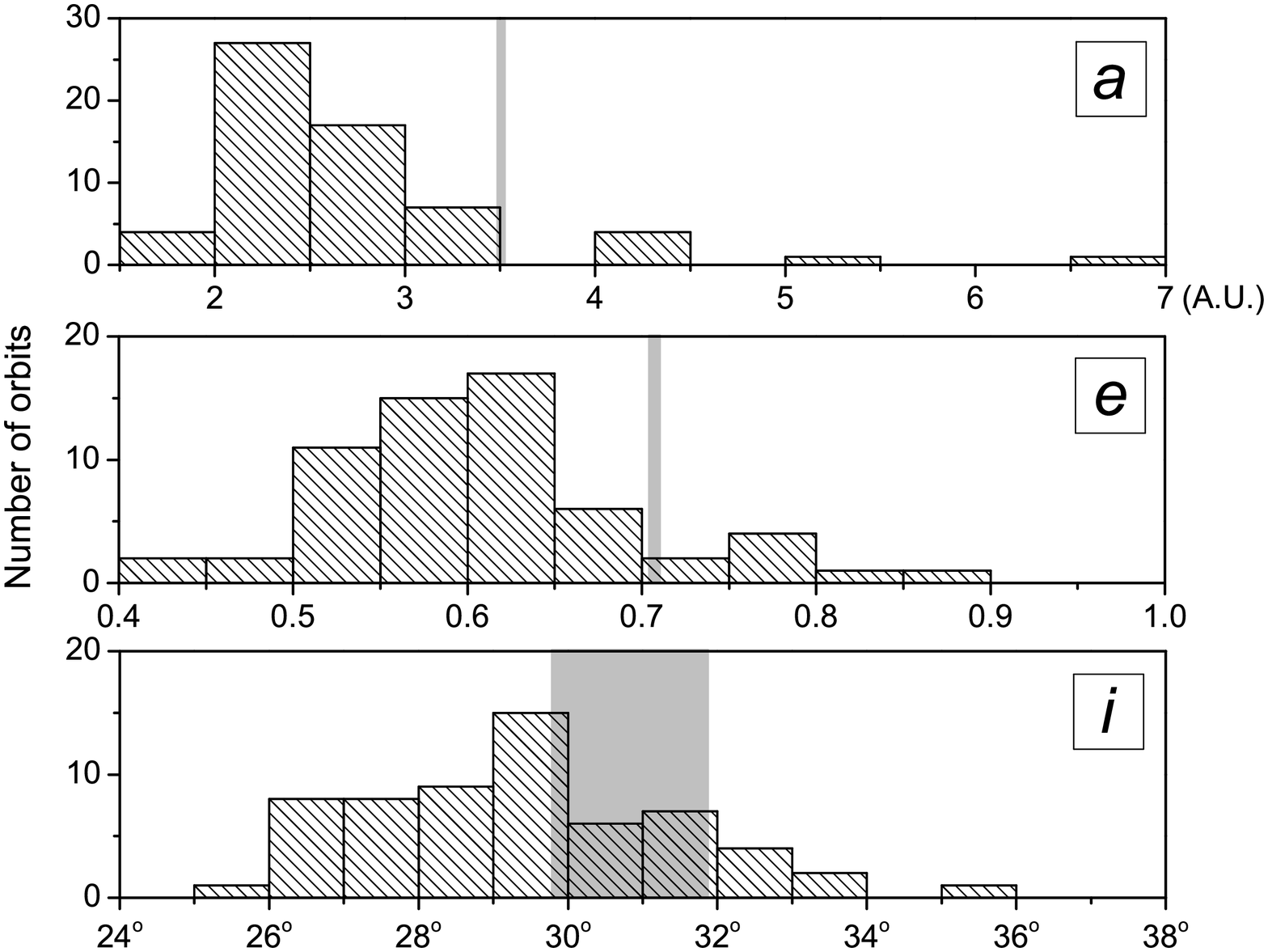}
\caption{The distribution of orbital elements $a$, $e$ and $i$ for the \texttt{complete} dataset. Shaded areas depict the range of respective orbital elements of the pseudo particles in mass range of $10^{-10}$ to $10^{-1}~\mathrm{kg}$ as simulated by \citet{vau11} (assuming bulk density of $300~\mathrm{kg \cdot m^{-3}}$).}
\label{orb-all}
\end{figure}

\begin{figure}
\includegraphics[scale=.3]{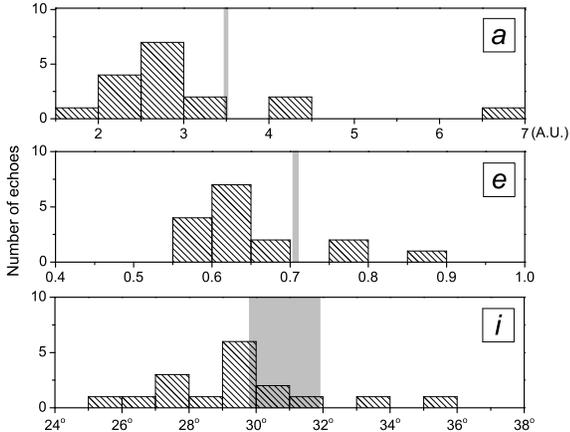}
\caption{The distribution of orbital elements $a$, $e$ and $i$ for the echoes in the \texttt{complete} dataset which $\Delta a/a<0.2$. Shaded areas depict the range of respective orbital elements of the pseudo particles in mass range of $10^{-10}$ to $10^{-1}~\mathrm{kg}$ as simulated by \citet{vau11} (assuming bulk density of $300~\mathrm{kg \cdot m^{-3}}$).}
\label{orb-best}
\end{figure}

The multi-station observations in the \verb"complete" dataset also allow determination of the orbits. The distribution of $a$, $e$ and $i$ is shown as Figure~\ref{orb-all}. We also selected a set of echoes with $\Delta a/a$ is below 0.2 ($\Delta a$ was determined by the echo simulator as described in \S3), defining a set of high quality multi-station orbits. The distribution of these orbits is shown in Figure~\ref{orb-best}. We also plot the range of $a$, $e$ and $i$ of the simulated particles by \citet{vau11} as shaded bars in each graph. It can be seen that most echoes are located in $a\in(2,3)$ which is smaller than the $a\sim3.5$ of the simulated particles, as well as the parent body.

\begin{figure}
\includegraphics[scale=.3]{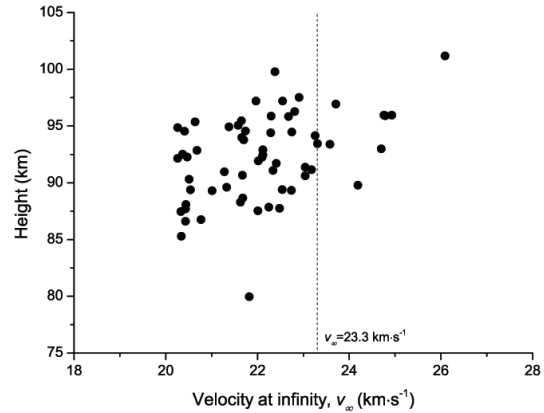}
\caption{Height versus deceleration corrected apparent velocity determined using the \texttt{tof} method corrected for mean deceleration ($v_{\infty}$) for all detected echoes in the \texttt{complete} dataset. If the deceleration correction was correct, the data points should not show any trend with height. The vertical line represents the theoretical deceleration corrected apparent velocity (23.3 $\mathrm{km\cdot s^{-1}}$) from \citet{jen06}}.
\label{ht-velc}
\end{figure}

What is the cause of this difference in $v_g$ and $a$ between observation and theoretical values? One possible explanation for the smaller observed $a$ (and hence smaller $v_g$) as compared to modeling, as noted by \citet{bro04} and \citet{cam06}, is an underestimation of the deceleration correction for Draconid echoes using the standard deceleration correction applied to CMOR echoes as a whole. Since the correction was derived from other showers which are not as fragile as the Draconids \citep[e.g.][]{fio64}, the out-of-atmosphere velocity ($v_{\infty}$) for Draconids may therefore be underestimated, resulting in a lower apparent initial velocity corresponding to a smaller $a$. As shown in Figure~\ref{ht-velc}, instead of the corrected $v_\infty$ showing no height trend, we observe a slowly increasing $v_{\infty}$ with respect to height, indicating a stronger deceleration to the meteoroids than the \citet{bro04} algorithm predicts.

\begin{figure}
\includegraphics[scale=.3]{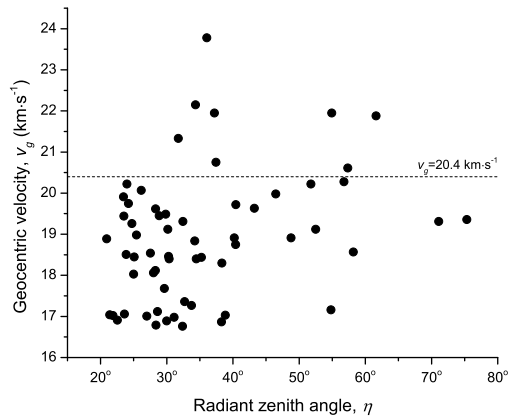}
\caption{Geocentric speed ($v_g$) versus radiant zenith angle ($\eta$) for all detected echoes in the \texttt{complete} dataset. A trend can be seen where $v_g$ is lower for smaller $\eta$, where trails were more vertical and were able to penetrate to lower heights.}
\label{vg-eta}
\end{figure}

\begin{figure}
\includegraphics[scale=.3]{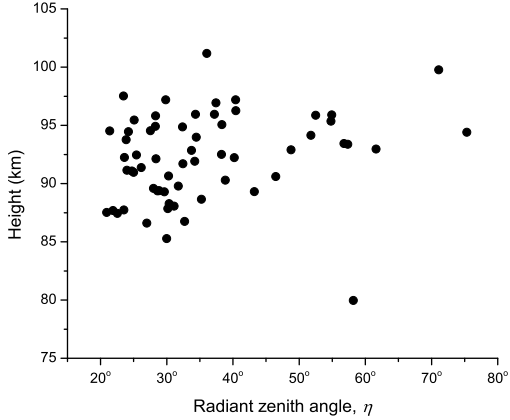}
\caption{Specular height versus radiant zenith angle ($\eta$) for all detected echoes in the \texttt{complete} dataset. A trend can be seen where the specular height is lower for smaller $\eta$, supporting the argument that trails that were more vertical were able to penetrate to lower heights.}
\label{ht-eta}
\end{figure}

However, from Table~\ref{tbl1} we also notice that the observed $v_g$ for Draconid echoes occurred during the times of the expected arrival of the 1873--1894 and 1900--1907 trails are even lower than the overall $v_g$. A quick data check shows that the echoes detected during the passage of the two trails were roughly 10--15$\%$ slower than the those detected in the early hours of Oct 8 (the weighted mean of echoes detected before 12h UT is $v_g=19.6\pm0.7~\mathrm{km \cdot s^{-1}}$). Although a difference in physical properties between these two trails and the Draconid background is a possible explanation, we suggest that the contribution of a small radiant zenith angle ($\eta$) (Figure~\ref{vg-eta}) is more convincing. Due to a higher radiant elevation, the meteoroid trajectory is more vertical, resulting in a deeper penetration through the atmosphere \citep[Figure~\ref{ht-eta}; see also][Equation 25]{ver73}. We note that the $v_g$ for 1900--1907 trails derived from European video observations
agrees the expected value \citep[$v_g=20.9\pm1.0~\mathrm{km \cdot s^{-1}}$ as given by][]{jen11} which supports our argument that the CMOR speeds are lower due to underestimation in deceleration.

\subsection{Mass Distribution}

One of the diagnostic parameters one can get from statistical analysis of a meteor outburst is the mass distribution index $s$. The mass index is a measure of mass distribution of the meteoroids; it is defined such that the number of meteoroids between mass interval $m$ and $m+dm$ is a power law described by $m^{-s}$.

It is very difficult from specular radar echoes to directly measure $m$ for each meteoroid; however, in radar theory, $m$ and the electron line density $q$ are linearly related under certain assumptions \citep[e.g.][\S7.3-7.5]{mck61}. Considering that $q$ is proportional to echo amplitude for underdense echoes, we can simply substitute amplitude for $m$, and use the slope of cumulative logarithmic amplitude distribution to derive $s$ \citep{mci68,sim68}. For overdense echoes, a diffusion-limited echo duration implies $\tau \propto q$ \citep[][\S8.9]{mck61}, and we can use echo duration instead of amplitude to estimate $s$ \citep{mci68} independently.

We have few underdense echoes due to the poor radiant geometry at the outburst peak for CMOR, therefore we first use the echo duration method to estimate $s$. However, since overdense echoes are much more likely to be affected by winds (Figure~\ref{fig-echo-sample}; which do not affect very short-lived underdense echoes), and are more prone to secondary specular point development \citep[see][\S4.9 and \S8.11]{mck61} and the radiant may not be properly determined, leading to possible sporadic contamination. To address this issue, we manually inspected every echo for its pre-t$_0$ phase feature (Figure~\ref{fig-pret0}). Only specular echoes exhibit such a feature, and therefore we can remove non-specular echoes. In the end, we selected a total of 155 Draconid overdense echoes on Oct. 8 in the \verb"overdense" dataset.

\begin{figure}
\includegraphics[scale=.3]{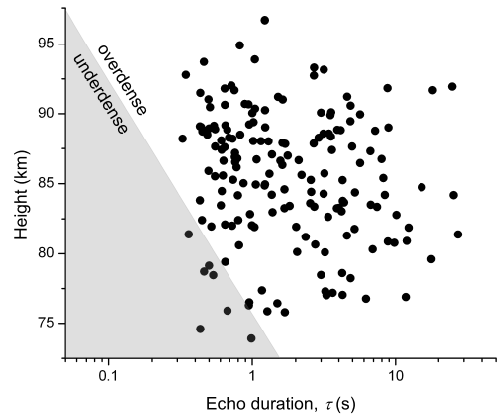}
\caption{The height range of the selected overdense echoes in \texttt{overdense} database. It can be seen that the lower limit of echo height is $\sim75~\mathrm{km}$. The shaded area marks the underdense region defined by Equation 1. We note that if this population were mainly underdense, a clear duration vs. height trend would be present.}
\label{tc-height}
\end{figure}

The identification of an overdense echo is rather subjective, leading to doubt as to whether some underdense echoes may have been confused with short-duration overdense echoes. However, unlike the duration of overdense echoes, which depends on radar wavelength, ambipolar diffusion coefficient, and electron line density, the duration of underdense echoes only depends on the radar wavelength and ambipolar diffusion coefficient (and hence height), as shown in \citet[][\S8.5]{mck61}. Combining this with the empirical diffusion coefficient formula derived by \citet{mas71}, we have

\begin{equation}
t_{ud} = \frac{\lambda^2}{16 \cdot \pi^2 \cdot 10^{0.06h-4.74}}
\end{equation}

where $t_{ud}$ is the duration and $\lambda$ is the radar wavelength, both in SI units, and $h$ is the height in kilometers. With $\lambda\sim10~\mathrm{m}$ and a lower limit of $h\sim75~\mathrm{km}$ (Figure~\ref{tc-height}), we see the upper limit of $t_{ud}$ to be $\sim 1~\mathrm{s}$. Based on this result, it is reasonable to label any CMOR echo with duration longer than $\sim 1~\mathrm{s}$ as an overdense echo.

\begin{figure}
\includegraphics[scale=.3]{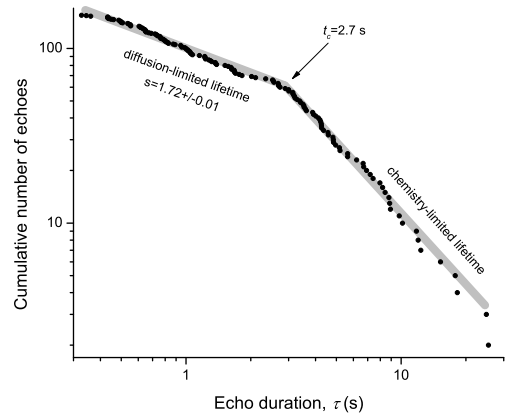}
\caption{The mass index of Draconids on Oct. 8 determined with selected echoes in the \texttt{overdense} dataset. The turnover from diffusion-limited to chemistry-limited duration (the duration of meteor echo limited by dissociating recombination between meteoric ions and atmospheric ozone molecules, \citet[c.f.][]{jon90}) can be seen at $t_c=2.7~\mathrm{s}$. The mass distribution index determined by fitting the diffusion-limited portion is $1.72 \pm 0.01$.}
\label{over-massindx}
\end{figure}

We found $s=1.72 \pm 0.01$ by fitting the linear portion of diffusion-limited overdense duration data points with Figure~\ref{over-massindx} and using the relation $N \propto \tau^{3(1-s)/4}$ \citep{wei61}, where $N$ is the cumulative number of echoes. The uncertainty here only depicts the uncertainty of least-squares fit.

\begin{figure}
\includegraphics[scale=.25]{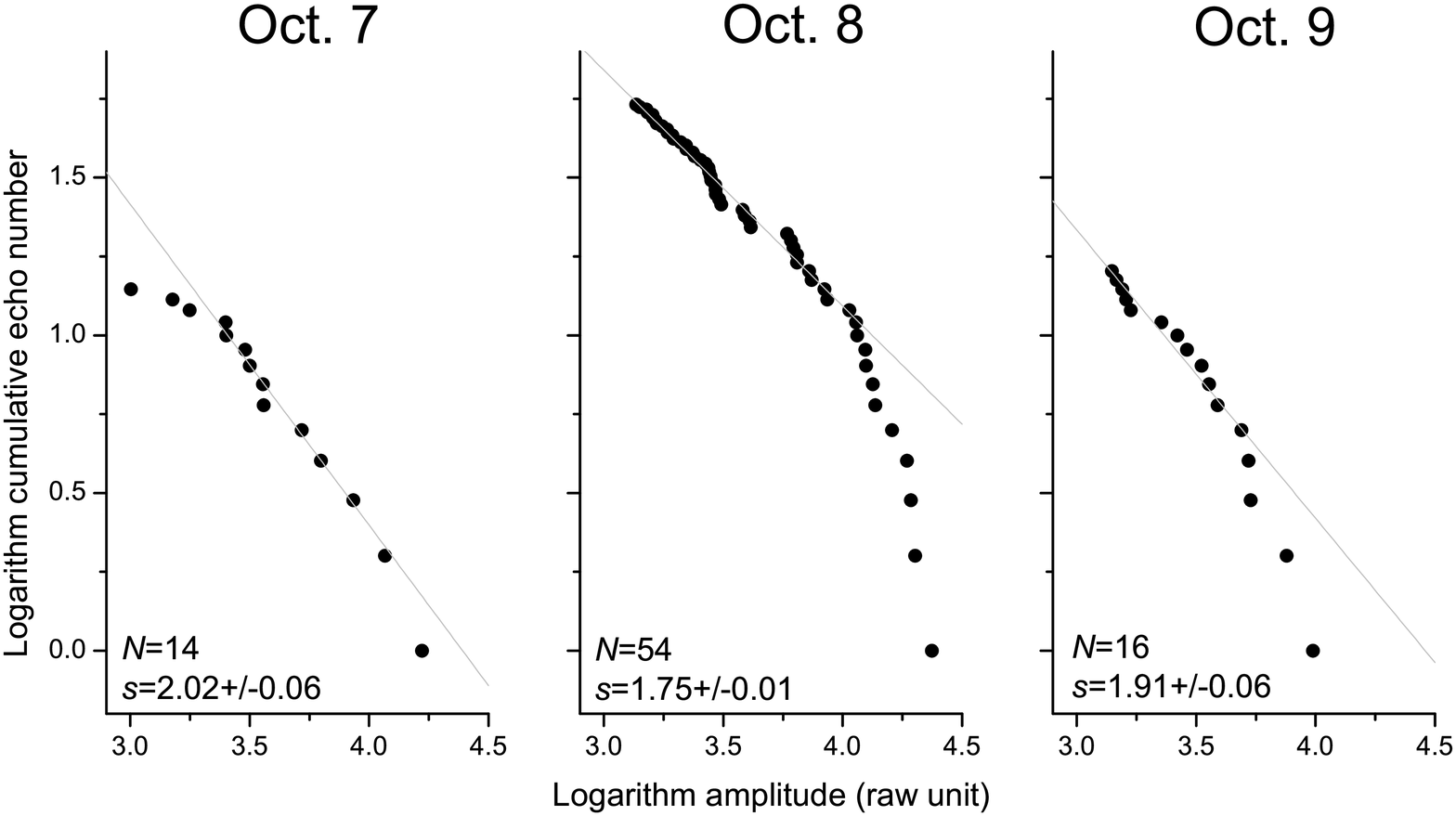}
\caption{Mass indices for Draconid underdense echoes between Oct. 7--9 determined using the underdense amplitude method.}
\label{mevhuge-massindx}
\end{figure}

We can also measure $s$ using the underdense echo amplitudes detected at the main site (Zehr). The advantage of this method is that it will include more events, including the possible Draconids that were filtered out in multi-station trajectory measurements. Sporadic contamination is possible, but given the low background rates at this time of the day ($\sim 15$h local time), we expect it to be representative of the mass index of the outburst. We select all the echoes perpendicular to within $5^{\circ}$ of the apparent radiant and $v_g=20.4~\mathrm{km \cdot s^{-1}}$ (derived using rise-time estimation which can be applied to single station data) over a range of 10$\%$. We find 84 single station echoes match our acceptance criteria for the period of Oct. 7--9, and the mass index is determined by using the relation $N \propto A^{1-s}$ \citep{mci68} (where $N$ is the cumulative number of echoes exceeds amplitude $A$). As shown in Figure~\ref{mevhuge-massindx}, the mass index determined in this way for Oct. 8
was $s=1.75 \pm 0.01$. The uncertainty given here is from least-square fitting only and is several times smaller than the real uncertainty given the small sample size \citep[see][for discussion]{bla11}. The mass indices for Oct. 7 and 9 appear to be higher than Oct. 8, possibly suggesting sporadic contamination, but they are subject to low confidence since the sample sizes are too small for quantitative estimates of $s$ (14 for Oct. 7 and 16 for Oct. 9). We are unable to compare this with the mass indices determined from overdense echoes for those two days, as there are also very few Draconid overdense echoes ($<10$) for either of the days.

The $s=1.75$ on Oct. 8, the peak, is low compared to that found by \citet{sim94} ($s=2.06$ for underdense echoes, $s=2.11$ for overdense echoes) for the 1985 Draconid outburst. The implies that the 2011 outburst was richer in bright meteors than the 1985 outburst. Of course, such a conclusion is weakened by the small sample size that we used, but we note that it is consistent with the theoretical suggestion by Maslov\footnote{\url{http://feraj.narod.ru/Radiants/Predictions/1901-2100eng/Draconids1901-2100predeng.html} (accessed Mar. 17, 2013).}, whose model suggested that the 1985 outburst was composed of ``quite faint meteors'' while the 2011 outburst was relatively rich in bright meteors. Combining the value derived from \verb"overdense" and single station data, we suggest the 2011 outburst had $s\sim1.75$, appropriate to $+3 \leq M_V \leq +7$.

\subsection{Flux}

\begin{figure*}
\includegraphics[scale=.5]{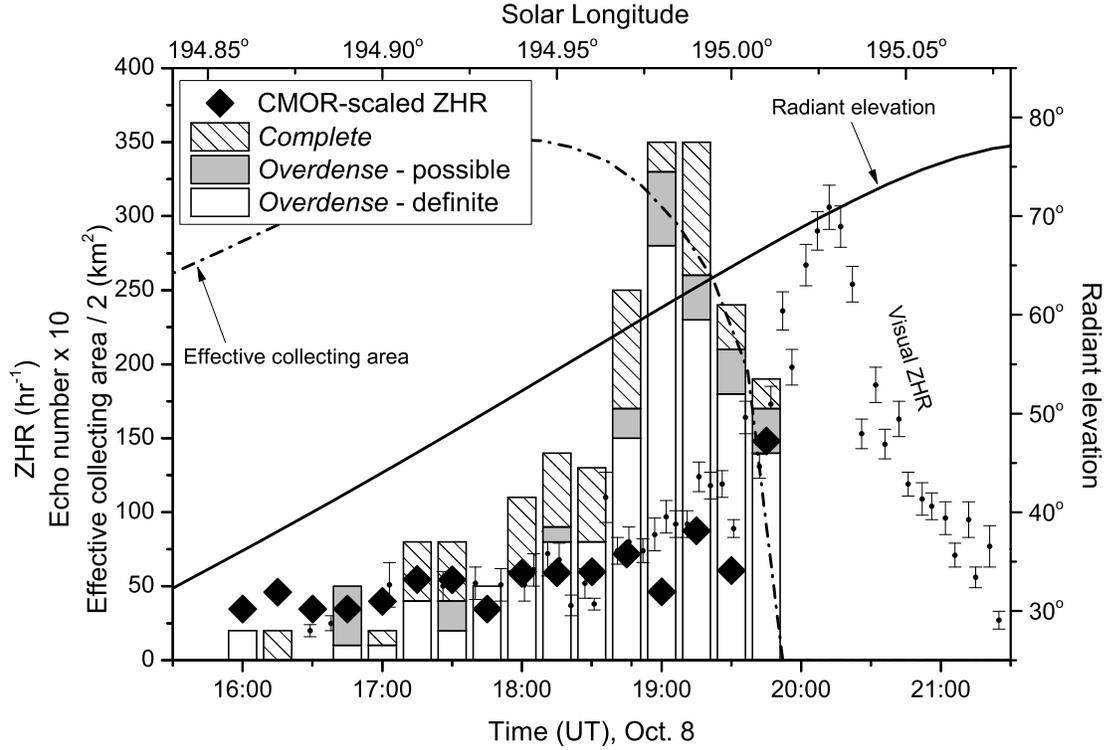}
\caption{The flux variation of Draconid outburst on Oct. 8 as observed by CMOR. \textit{Complete} stands for the number of echoes in the \texttt{complete} dataset, \textit{overdense - possible} stands for the possible Draconid overdense echoes in the \texttt{overdense} dataset, and \textit{overdense - definite} stands for the definite Draconid overdense echoes in the \texttt{overdense} dataset (see \S3.1 for details). The overdense data is scaled relative to the visual data from IMO, and plotted on the same scale.}
\label{flux}
\end{figure*}

We counted the events in the \verb"overdense" dataset and selected events around the time of the maximum in the effective specular collecting area (17:30--18:30 UT) to compare to the zenith hourly rate (ZHR) published by the International Meteor Organization\footnote{\url{http://www.imo.net/live/draconids2011/} (accessed Mar. 17, 2013).}, shown in Figure~\ref{flux}. As noted in previous sections, the main peak measured with visual data for the 2011 outburst occurred under very poor specular scattering geometry for CMOR, such that only the outburst rise is measurable. However, the scaled ZHR from the overdense radar data shows a similar variation to the visual data.

\subsection{Meteoroid Structure Inferred from Ablation Modeling}

The meteoroid ablation model developed by \citet{cam04} was used to explore the structure of radar-detected Draconid meteoroids. We can use the velocities and electron line densities measured from different radar stations as constraints, and fit the model to match the observations. The velocity used here is the \texttt{tof} velocity described in \S3; the electron line density is computed from the amplitude/power of the echo \citep[see][\S4.2, \S4.3]{cep98}. The computation of electron line density differs between underdense and overdense echoes. Following the definition of classic radar meteor theory, we use the underdense formula for echoes with $q<2.4\times 10^{14}~\mathrm{m^{-1}}$ and vice versa \citep[][Equation 8-29]{mck61}. In this way, we perform the ablation entry modeling using two approaches: (i) modeling the Draconids as a ``mean'' population using all events; (ii) modeling individual events that showed noticeable deceleration across multiple stations.

\subsubsection{Modeling Draconids as a Population}

In this approach, we mainly focused on two meteoroid structural parameters: grain number ($n_{grain}$) and grain mass ($m_{grain}$); for other tunable parameters, we either use the known properties of Draconids or those used by \citet{cam04}, as summarized in Table~\ref{tbl-amodel}.

\begin{table}
 \centering
  \caption{Fixed input parameters of the \citet{cam04} model. The deceleration corrected apparent velocity $v_{\infty}$ is from \citet{jen06}; the zenith angle uses the most probable value from our data; the bulk and grain densities use the values suggested by \citet{bor07}; the heat of ablation and thermal conductivity use the default values used by \citet{cam04}.}
  \begin{tabular}{cc}
  \hline
  Parameter & Value \\
  \hline
Deceleration corrected apparent velocity, $v_{\infty}$ & 23.27 $\mathrm{km \cdot s^{-1}}$ \\
Zenith angle, $\eta$ & $40^{\circ}$ \\
Bulk density, $\rho_{bulk}$ & 300 $\mathrm{kg \cdot m^{-3}}$ \\
Grain density, $\rho_{grain}$ & 3 000 $\mathrm{kg \cdot m^{-3}}$ \\
Heat of ablation, $q$ & $3 \times 10^6$ $\mathrm{J \cdot kg^{-1}}$ \\
Thermal conductivity, $\kappa$ & 0.2 $\mathrm{J \cdot m^{-1} \cdot s^{-1} \cdot K^{-1}}$ \\
\hline
\end{tabular}
\label{tbl-amodel}
\end{table}

\begin{figure*}
\includegraphics[scale=.8]{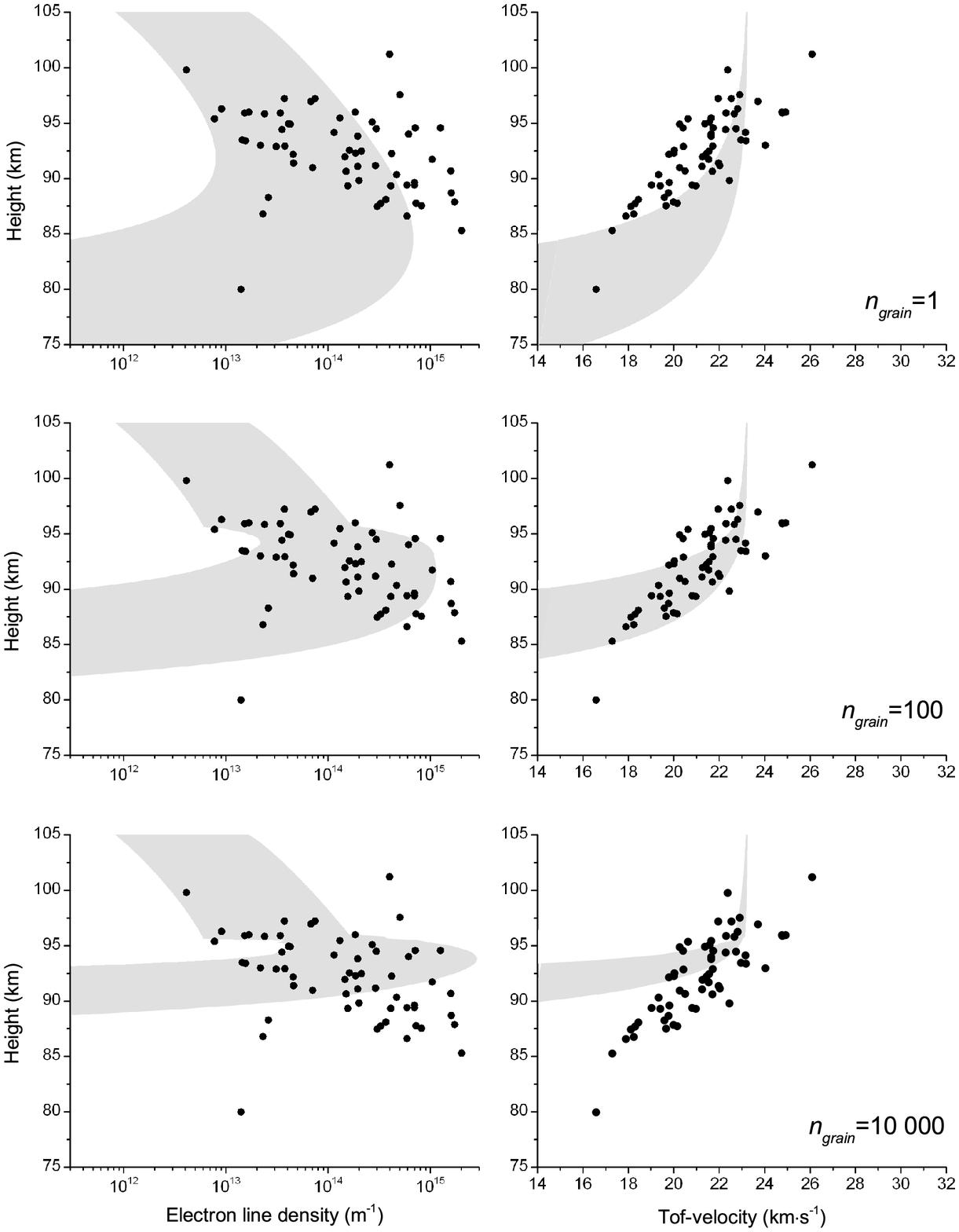}
\caption{Modeling results of three ($m_{grain}, n_{grain}$) combinations together with radar observations. The total meteoroid mass is fixed at the range of [$10^{-6}~\mathrm{kg}$, $10^{-4}~\mathrm{kg}$], while the combination of ($m_{grain}, n_{grain}$) varies from ([$10^{-6}~\mathrm{kg}$, $10^{-4}~\mathrm{kg}$], 1) for the upper set, ([$10^{-8}~\mathrm{kg}$, $10^{-6}~\mathrm{kg}$], 100) for the middle set, and ([$10^{-10}~\mathrm{kg}$, $10^{-8}~\mathrm{kg}$], 10 000) for the lower set. The curves stand for the modeling result using certain ($m_{grain}$, $n_{grain}$). It can be seen that the best fit is the middle set, with $m_{grain} \in [10^{-8}~\mathrm{kg}$, $10^{-6}~\mathrm{kg}$] and $n_{grain}=100$.}
\label{frag-test}
\end{figure*}

First, we find starting values for ($m_{grain}, n_{grain}$). We fix the total meteoroid mass in the range [$10^{-6}~\mathrm{kg}$, $10^{-4}~\mathrm{kg}$], and plot the results of three combinations with ($m_{grain}, n_{grain}$) ranging from $(10^{-4}~\mathrm{kg},1)$ to $(10^{-10}~\mathrm{kg},10~000)$ to compare with radar observations, as shown in Figure~\ref{frag-test}. Our best fit is near $m_{grain} \in [10^{-8}~\mathrm{kg},~10^{-6}~\mathrm{kg}]$ and $n_{grain}=100$, though the solution is not unique and we emphasize that we are in effect trying to fit an average Draconid, when there may be significant variation within the stream.

\begin{figure}
\includegraphics[scale=.4]{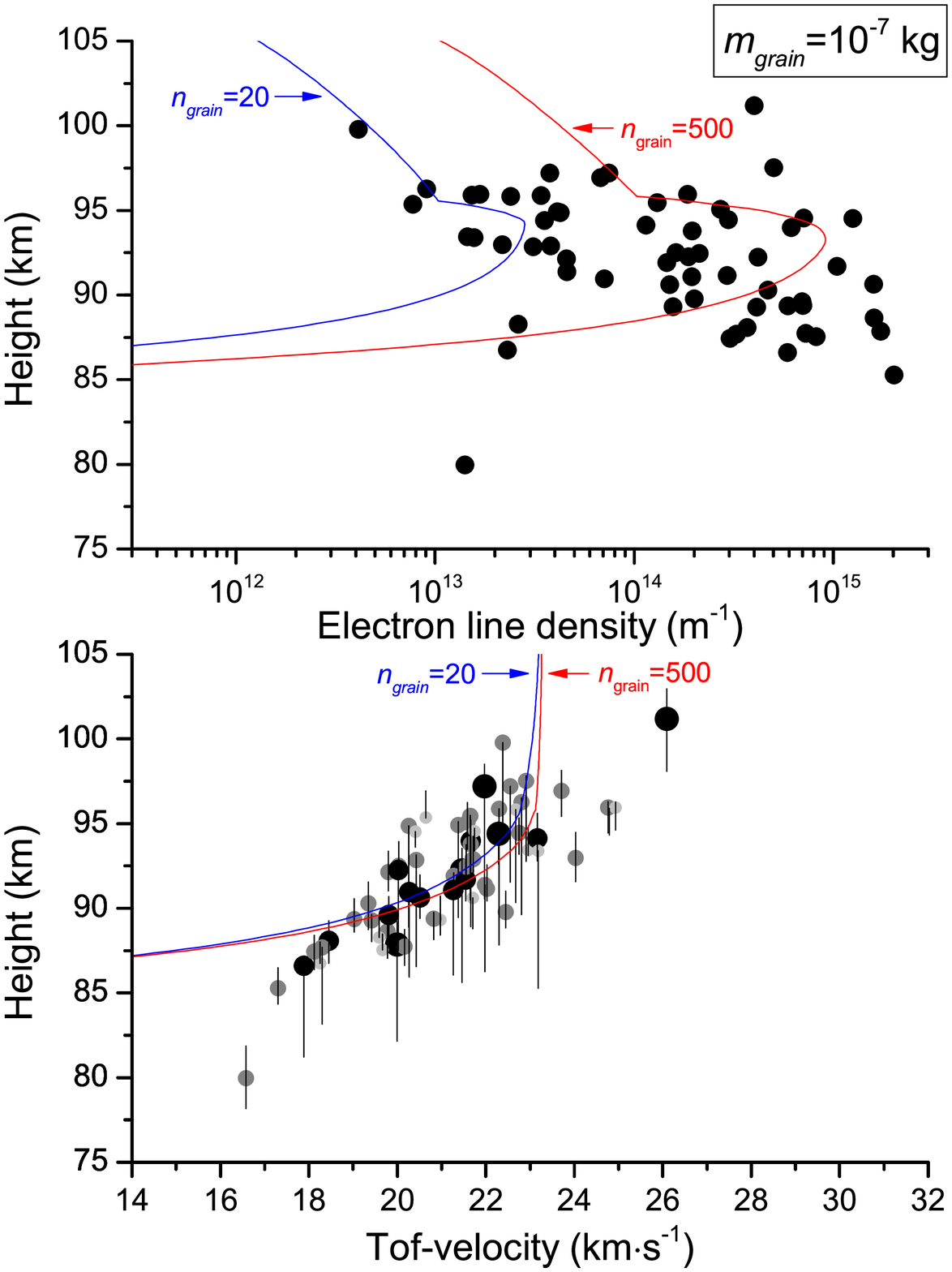}
\caption{Modeling results with fixed $m_{grain}=10^{-7}~\mathrm{kg}$ and variable $n_{grain}=20$, $500$, corresponding to total masses of $2 \times 10^{-6}~\mathrm{kg}$ and $5 \times 10^{-5}~\mathrm{kg}$. Size and color of data points in velocity figure indicates number of sites: darker and larger points indicates more sites and vice versa. The vertical line for each data point depicts the height range constrained by multi-site observations. Curves depict modeling solutions under different $n_{grain}$. The modelled fit does not show as much scatter as the observations compared with Figure~\ref{frag-n-fixed}.}
\label{frag-m-fixed}
\end{figure}

\begin{figure}
\includegraphics[scale=.4]{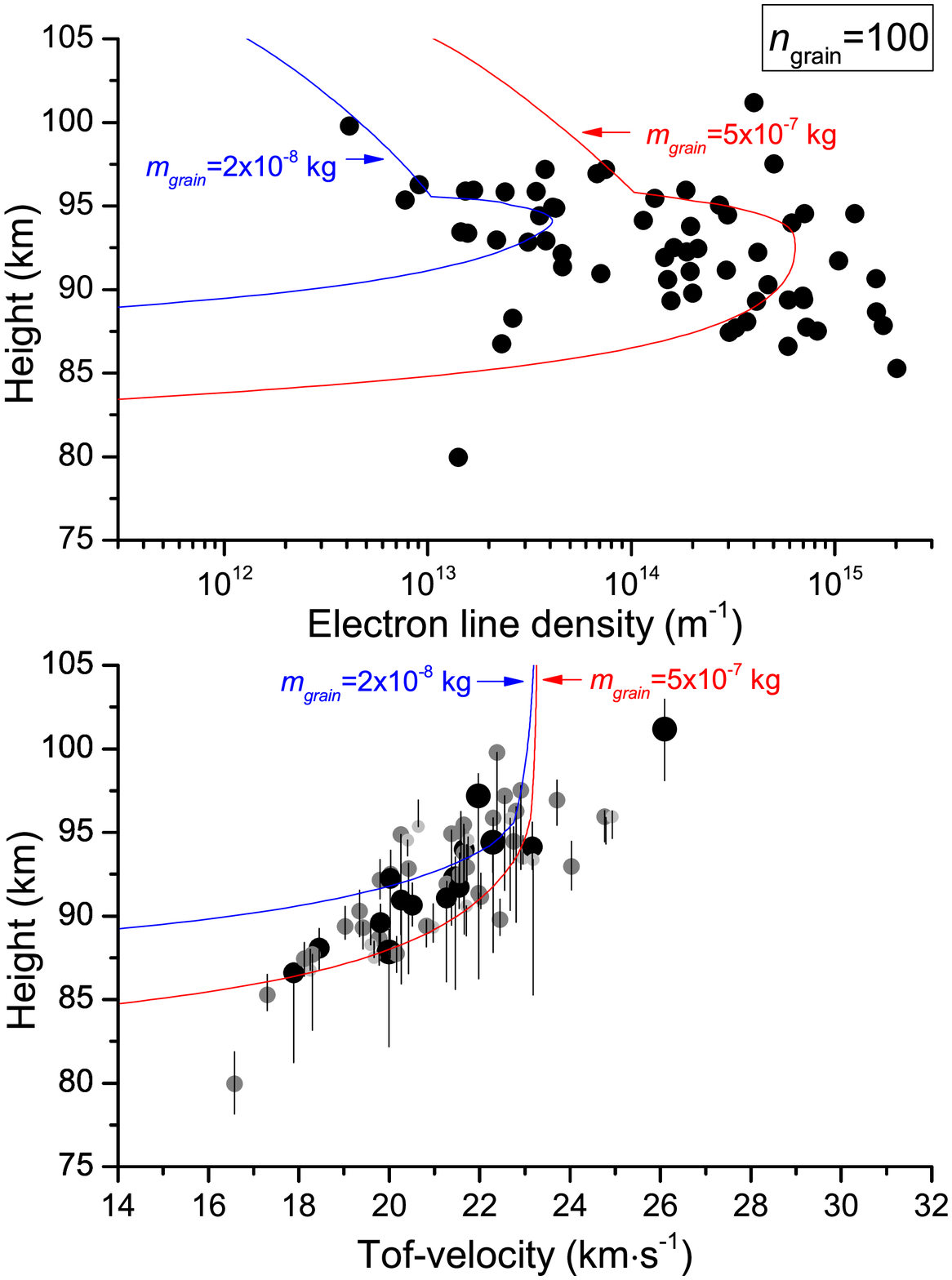}
\caption{Modeling results with fixed $n_{grain}=100$ and variable $m_{grain}=5 \times 10^{-7}~\mathrm{kg}$, $2\times 10^{-8}~\mathrm{kg}$, corresponding to total masses of $2 \times 10^{-6}~\mathrm{kg}$ and $5 \times 10^{-5}~\mathrm{kg}$. Size and color of data points in velocity figure indicates number of sites: darker and larger points indicates more sites and vice versa. The vertical line for each data point depicts the height range constrained by multi-site observations. Curves depict modeling solutions under different $n_{grain}$. Comparing with Figure~\ref{frag-m-fixed}, we see a better match to the observations especially at middle and lower heights.}
\label{frag-n-fixed}
\end{figure}

Next we want to see which of the two parameters ($m_{grain}, n_{grain}$) dominates the process. We investigate this by fixing either $m_{grain}$ (to $10^{-7}~\mathrm{kg}$) or $n_{grain}$ (to 100), letting the other parameter, as well as total meteoroid mass, being variable as summarized in Table~\ref{tbl-frag-fix}. The results are shown in Figure~\ref{frag-m-fixed} and Figure~\ref{frag-n-fixed}. For both the electron line density and velocity simulations, we see a trend that fixed-$n_{grain}$ curves match the observations better; for the fixed-$m_{grain}$ curves, the sharp maximum of electron line densities near 93 km does not match the observed data points over the same height range. Also the velocity simulation showed little variation with different $n_{grain}$ in contrast to the observation.

\begin{table}
 \centering
  \caption{Summary of parameters for the sensitivity test of $m_{grain}$ and $n_{grain}$ for Figure~\ref{frag-m-fixed} and Figure~\ref{frag-n-fixed}.}
  \begin{tabular}{cccc}
  \hline
  Figure & $m_{grain}$ (average) & $n_{grain}$ & Total mass \\
  \hline
Figure~\ref{frag-m-fixed} & $10^{-7}~\mathrm{kg}$ (fixed) & 20 & $2 \times 10^{-6}~\mathrm{kg}$ \\
 & & 500 & $5 \times 10^{-5}~\mathrm{kg}$ \\
  \hline
Figure~\ref{frag-n-fixed} & $2 \times 10^{-8}~\mathrm{kg}$ & 100 (fixed) & $2 \times 10^{-6}~\mathrm{kg}$ \\
& $5 \times 10^{-7}~\mathrm{kg}$ & & $5 \times 10^{-5}~\mathrm{kg}$ \\
\hline
\end{tabular}
\label{tbl-frag-fix}
\end{table}

\subsubsection{Modeling Individual Draconid Echoes}

\begin{figure}
\includegraphics[scale=.3]{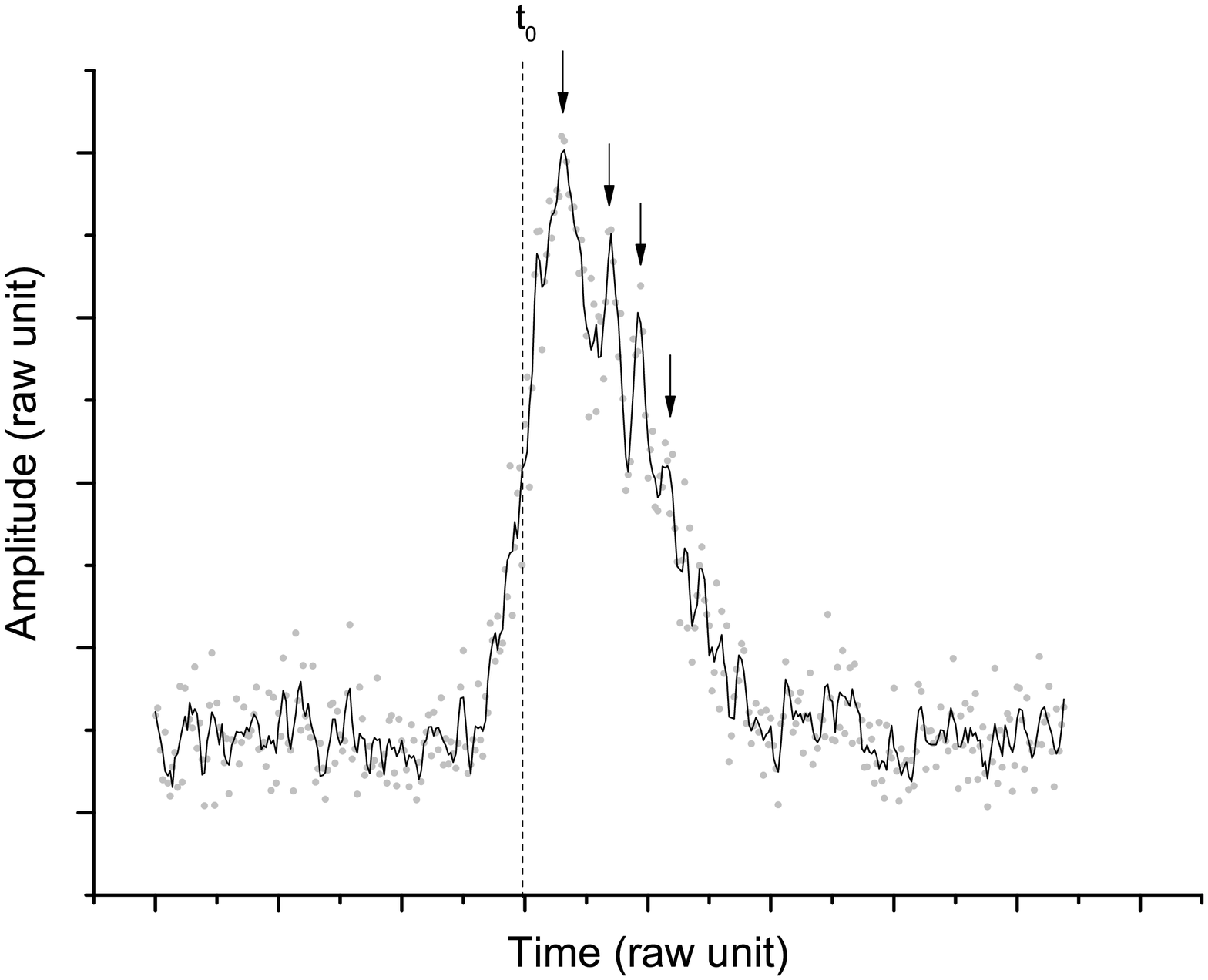}
\caption{Example of an underdense Draconid echo with four Fresnel maxima amplitude features noted (marked by arrows).}
\label{fresnel}
\end{figure}

\begin{figure}
\includegraphics[scale=.3]{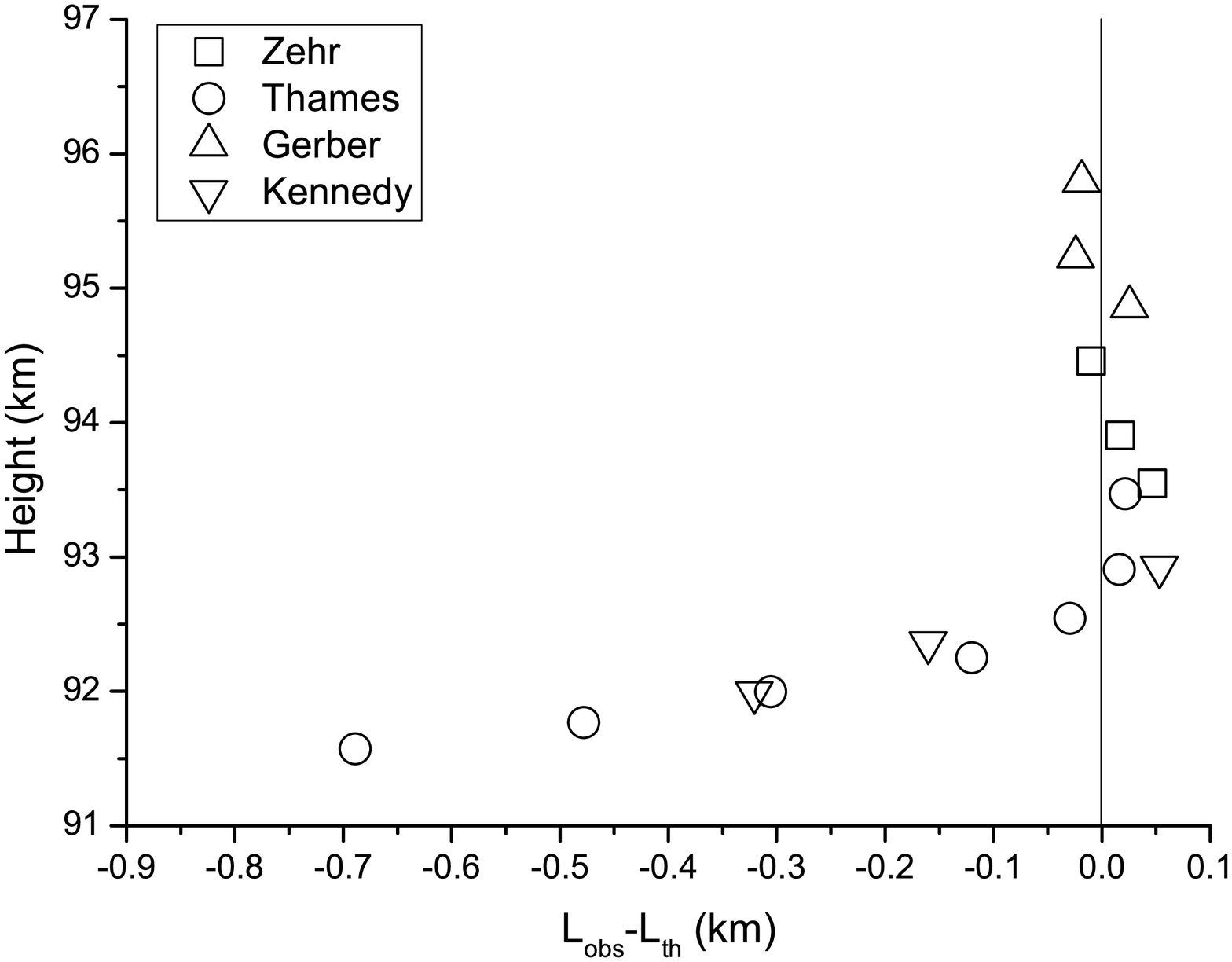}
\caption{The deceleration of the 17:31:47 event, plotted in position offset versus height. We linearly extrapolate the linear portion of the data points (from a height of 96 to 92.5 km), the position offset is the distance between the observed position and the linear extrapolation.}
\label{decobs}
\end{figure}

As the time and phase measurements in CMOR data are synchronized, it is possible to use the Fresnel amplitude fluctuation features just after the $t_0$ point (Figure~\ref{fresnel}) to precisely measure the positions \citep[see][\S4.6.1 and references therein]{cep98}; we therefore searched in the \texttt{complete} dataset for any echo showing significant deceleration using this method. We have many echoes which show no measurable deceleration, but this does not provide a strong constraint on the entry model parameters, so we focus instead on events with clear deceleration. Since the same meteor seen at different sites usually corresponds to different positions on the trail, if the Fresnel feature shows up in observations from more than one site, we can obtain a series of contiguous position measurements along the trail. If there is no deceleration, then the position should increment linearly with time; therefore, by plotting position-time series, we can readily see if there is any deceleration. This is
analogous to the distance along trail versus time plots used by \citet{bor07} for video data to reveal meteoroid deceleration. In this way, we identified one event (detected on 17:31:47 UT on Oct. 8 at the main site) that showed significant deceleration (Figure~\ref{decobs}). No other Draconid showed measurable deceleration.

\begin{figure}
\includegraphics[scale=.4]{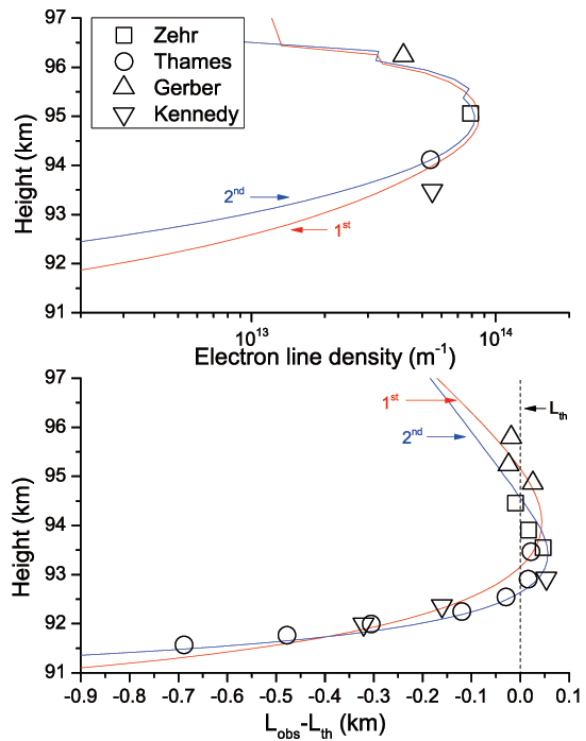}
\caption{The modeling result for the 17:31:47 event: upper plot -- modeling and observed electron line density; lower plot -- modeling and observed theoretical minus observational position assuming a fixed speed. The theoretical fit was extrapolated using the position from 96 to 92.5 km as reference.}
\label{decmod}
\end{figure}

To model this event, we start from the same set of parameters given in Table~\ref{tbl-amodel} except the zenith angle and entry velocity are based on the specific echo observation (measured to be $36.8^{\circ}$ and $24.3~\mathrm{km \cdot s^{-1}}$ from the specular position and pre-t$_0$ velocity measured at the main site). We take two approaches: for the first, we fix the bulk and grain density to the one recommended by \citet{bor07} (i.e. the one used in the previous section, $\rho_{bulk} = 300~\mathrm{kg \cdot m^{-3}}$ and $\rho_{grain} = 3~000~\mathrm{kg \cdot m^{-3}}$), then tune $m_{grain}$ and $n_{grain}$ until a minimum difference between the modeling and the observation is achieved. The second approach leaves both the bulk and the grain density as well as the zenith angle and entry velocity flexible to eliminate any remaining differences. The results found with the two modeling approaches, as well as the input parameters, are given in Figure~\ref{decmod} and Table~\ref{tbl-amodel-dec}.

\begin{table*}
 \centering
  \caption{Input parameters of two approaches used to model the ablation of the 17:31:47 event. For the 1$^\mathrm{st}$ approach, the bulk and grain densities are set to tose given by \citet{bor07}, while for the 2$^\mathrm{nd}$ approach the two are fit by the model.}
  \begin{tabular}{ccc}
  \hline
  Parameter & 1$^\mathrm{st}$ approach & 2$^\mathrm{nd}$ approach \\
  \hline
Number of grains, $n_{grain}$ & 800 & 4 000 \\
Grain mass (average), $m_{grain}$ (kg) & $3\times10^{-9}$ & $5\times10^{-10}$ \\
Grain mass range (kg) & [$10^{-9}$,$4\times10^{-9}$] & [$3\times10^{-11}$,$5\times10^{-10}$] \\
Total mass, $m$ (kg) & $2.4\times10^{-6}$ & $2\times10^{-6}$ \\
Deceleration corrected apparent velocity, $v_{\infty}$ ($\mathrm{km \cdot s^{-1}}$) & 24.30 & 24.50 \\
Zenith angle, $\eta$ & $38.3^{\circ}$ & $36.8^{\circ}$ \\
Bulk density, $\rho_{bulk}$ ($\mathrm{kg \cdot m^{-3}}$) & 300 & 500 \\
Grain density, $\rho_{grain}$ ($\mathrm{kg \cdot m^{-3}}$) & 3 000 & 1 500 \\
\hline
\end{tabular}
\label{tbl-amodel-dec}
\end{table*}

\subsubsection{Comparison and Discussion}

We compare our observational and modeling result to \citet{bor07}, who reported six decelerated Draconid meteors with sizes comparable to radar meteors, detected by double-station video system during the 2005 Draconid outburst. Their grain masses were in the range of $10^{-11}$ to $10^{-9}~\mathrm{kg}$, with $\rho_{bulk}=300~\mathrm{kg \cdot m^{-3}}$ with the assumption of $\rho_{grain}=3~000~\mathrm{kg \cdot m^{-3}}$, and total number of grains between $10^4$ and $10^6$. These six meteors all showed significant deceleration with turnover height (i.e. the height that $\dot{v}$ becomes significant) around 95 km. The differences with our results include a much smaller grain number ($10^2$) and a much larger grain mass (at the order of $10^{-7}~\mathrm{kg}$) despite the density configurations being similar. Our 17:31:47 event also showed a slightly lower turnover height ($\sim 93$ km) than the video meteors.

How to explain these differences? One possibility is the outbursts from 2005 and 2011 were ejected by Comet Giacobini-Zinner in different years: the material encountered in 2011 was ejected earlier than 2005 \citep[1873--1907 versus 1946, per][]{cam06,vau11}. However, we note that it is still difficult at this stage to uniquely define the physical properties of meteoroids in this way due to the lack of observational constraints. For example, the statistical model fits shown in Figure~\ref{frag-test} are broad averages which allow much uncertainties; for the modeling of the 17:31:47 event we have shown that a reasonable fit can be found with different sets of parameters, and similar phenomenon have been noted by the Borovi\v{c}ka et al. (see \S7.1 of their paper). All these factors cast doubts on whether we are indeed seeing the physical differences between meteoroids or just fit variations admittedable within the uncertainties. However, we still see that it is possible to use the observations to broadly
constrain meteoroid properties to some extent, and we feel encouraged by the fact that our results have shown basic agreement to the earlier work of \citet{bor07}.

Also, we do not have sufficient constrains on the chemical state of the meteor with radar observations, which prevent us from considering more chemistry-enhanced model, such as the differential ablation model developed by \citet{von08}. Future multi-instrumental observations, for example simultaneous radar-spectral observations, will be helpful to solve this problem and give more insights on the chemical state of the meteors.

\section{Conclusions}

In this study, we have analyzed CMOR observations of the 2011 Draconid outburst, including 61 multi-station specular echoes and 179 single-station overdense echoes. Our main results are:

\begin{enumerate}
  \item The radiant of the outburst was determined to be $\alpha_g=261.9^{\circ} \pm 0.3^{\circ}$, $\delta_g=+55.3^{\circ} \pm 0.3^{\circ}$ (J2000) with the multi-station echoes, which agreed with earlier modeling forecasts to $\sim1^{\circ}$.
  \item The averaged velocity determined from the CMOR data was by $\sim 10-15\%$ smaller than the expected value ($17.0-19.1$ $\mathrm{km \cdot s^{-1}}$ versus 20.4 $\mathrm{km \cdot s^{-1}}$), likely due to the undercorrection of deceleration, combined with the effect of radiant geometry during the peak hours;
  \item The mass distribution index determined from 155 overdense echoes was $s=1.72\pm0.01$ assuming the echoes used for such determination were diffusion-dominated. Alternatively, we selected 54 possible Draconids in the sample of single-station underdense echoes and determined the mass index to be around 1.75. Combining these values, we suggest that the mass index of the 2011 outburst to be $s=$1.75 for the magnitude range of $+3 \leq M_V \leq +7$;
  \item We compared the counts of overdense echoes to the visual data, and a clear consistency can be noted. Unfortunately, the peak time of the outburst occurred with very poor scattering geometry for CMOR, such that no comparison can be made at the peak time. However the rise of the meteor rates between the radar and visual results were highly comparable;
  \item We used the meteoroid ablation model developed by \citet{cam04} to explore the structure of decent radar Draconid meteoroids. Assuming bulk and grain densities to be $300~\mathrm{kg \cdot m^{-3}}$ and $3~000~\mathrm{kg \cdot m^{-3}}$ respectively, the model seems to infer that the grain number of CMOR-observed Draconid meteoroids is $\sim100$ regardless of meteoroid sizes;
  \item We also identified a Draconid meteor that showed clear deceleration. Two modeling approaches were attempted, suggested the grain numbers to be $\sim1~000$, grain mass between $10^{-10}$ to $10^{-9}$ kg, and the total mass to be around $2\times10^{-6}$ kg, which were in close agreement with the video results reported by \citet{bor07}.
\end{enumerate}

As this paper was being finished, an unexpected and much more intense Draconid outburst was detected by CMOR on Oct. 8, 2012. Preliminary analysis suggests a spectacular ZHR at the order of $8~000$, far above the storm level. In-depth investigation of the 2012 event will be addressed in a separate paper.

\section*{Acknowledgments}

We thank Zbigniew Krzeminski, Jason Gill and Daniel Wong for helping with CMOR analysis and operations, and J\'{e}r\'{e}mie Vaubaillon for providing additional numerical dynamical modelling data. We also thank an anonymous reviewer for comments. Funding support from the NASA Meteoroid Environment Office (cooperative agreement NNX11AB76A) for CMOR operations is gratefully acknowledged. QY thanks Xia Han, Abedin Y. Abedin and Reynold E. Sukara for valuable discussions. He also thanks the Comet 21P/Giacobini-Zinner for commitment to a slow suicide to make this study (and his financial support) possible.

Additionally, we thank the following observers for their reports during the IMO Draconid visual campaign: Ioan Agavriloaiei, Salvador Aguirre, Juan Carlos Alcazar Fernandez, Karl Antier, Rainer Arlt, Jure Atanackov, Luc Bastiaens, Zora Beljic, Orlando Ben\'{i}tez S\'{a}nchez, Felix Bettonvil, Suresh Bhattarai, Jens Briesemeister, St\'{e}phane Bruchet, the BSU students, Vladimir Burgic, Rafael Campillos Ladero, Jakub Cerny, Marcin Chwaa, Filip Colakovic, Lorenzo Comolli, Ilie Cosovanu, H\aa{a}kon Dahle, Anderson Dantas, Luigi D'argliano, C. B. Devgun Space, Jose Vicente Diaz Martinez, Sietse Dijkstra, Jan Ebr, Ivana Ebrova, Frank Enzlein, Tomasz Fajfer, Reyhaneh Falah, Richard Fleet, Stela Frencheva, Stefan Fuks, Kai Frode Gaarder, Sylvie Gorkova, Mitja Govedi, Matthias Growe, Piotr Guzik, Shy Halatzi, Oliver Hanke, Amir
Hasanzadeh, Leo Holmberg, Kamil Hornoch, Jan Horsky,
Antal Igaz, Jelisaveta Ilic, Nevena Ilic, Carl Johannink, Javor Kac, Timo Karhula, Roy Keeris, Stanislav Korotkiy, Janez Kos, Roman Kostenko, Jakub Koukal, Roman Kovalyk, Martin Krueger, Andrey Kychyzhyyev, Marina Kychyzhyyeva, Jens Lacorne, Marco Langbroek, Mariusz Lemiecha, Anna Levin, Anna Levina, Hartwig Luethen, Jose Luis Maestre Garcia, Alexandr Maidik, Maria Makarova, Veikko M\"{a}kel\"{a}, Alexander Manannikov, Dasha Maskovay, Aleksandar Matic, Ivo Micek, Marco Micheli, Maslov Mikhail, Jose Carlos Millan, Artem Mirgorod, Jakub Mirocha, Koen Miskotte, Sirko Molau, Maciej Myszkiewicz, Nikolay Nikolaev, Artyom Novichonok, Francisco Oc\~{n}a Gonz\'{a}lez, Peter I. Papics, Lovro Pavletic, Anna Pavlova, Konstantin Polyakov, Sasha Prokofyev, Rok Pucer, Jatin Rathod, Ella Ratz, Jurgen Rendtel, Raluca Rufu, Miguel Santana Guti\'{e}rrez, Krisztian Sarneczky, Mikiya Sato, Tomoko Sato, Branislav Savic, Christian Schmiel, Alex Scholten, Kai Schultze, Andrzej Skoczewski, Yaser Soleimani, Ulrich Sperberg, Israel
Tejera Falcon, Seityagiya Terlekchi, Snezana Todorovic, Yasuhiro Tonomura, Josep M. Trigo-rodr\'{i}guez, Blanca Troughton Luque, Shigeo Uchiyama, Michel Vandeputte, Wienie Van Der Oord, Peter Van Leuteren, Birgit Van Opstal, Glynis Van Uden, Jovan Vasiljevic, Marina Vlajnic, Marta Volkova, William Watson, Mariusz Wisniewski, A. O. Woost, Ilkka Yrj\"{o}l\"{a}, Raziyeh Zahedi, Weizhou Zeng, Tianwei Zhang, and Peter Zimnikoval.

\bibliographystyle{mn2e}
\bibliography{man}

\begin{thebibliography}{}

\bibitem[\protect\citeauthoryear{{Arlt}}{{Arlt}}{1998}]{arl98}
{Arlt} R.,  1998, WGN, Journal of the International Meteor Organization, 26,
  256

\bibitem[\protect\citeauthoryear{{Arlt}, {Bellot Rubio}, {Brown} \&
  {Gyssens}}{{Arlt} et~al.}{1999}]{arl99}
{Arlt} R.,  {Bellot Rubio} L.,  {Brown} P.,    {Gyssens} M.,  1999, WGN,
  Journal of the International Meteor Organization, 27, 286

\bibitem[\protect\citeauthoryear{{Asher}}{{Asher}}{1999}]{ash99}
{Asher} D.,  1999, in {Jenniskens} P.,  ed., The Leonid MAC Workshop {Modeling
  of the Leonid meteor shower}

\bibitem[\protect\citeauthoryear{{Baggaley}, {Bennett}, {Steel} \&
  {Taylor}}{{Baggaley} et~al.}{1994}]{bag94}
{Baggaley} W.~J.,  {Bennett} R.~G.~T.,  {Steel} D.~I.,    {Taylor} A.~D.,
  1994, \qjras, 35, 293

\bibitem[\protect\citeauthoryear{{Biot}}{{Biot}}{1841}]{bio48}
{Biot} E.~C.,  1841, {Catalogue general des etoiles filantes et des autres
  meteores observes EN Chine pendant vingt-quatre siecles}

\bibitem[\protect\citeauthoryear{{Blaauw}, {Campbell-Brown} \&
  {Weryk}}{{Blaauw} et~al.}{2011}]{bla11}
{Blaauw} R.~C.,  {Campbell-Brown} M.~D.,    {Weryk} R.~J.,  2011, \mnras, 412,
  2033

\bibitem[\protect\citeauthoryear{{Borovi{\v c}ka}, {Spurn{\'y}} \&
  {Koten}}{{Borovi{\v c}ka} et~al.}{2007}]{bor07}
{Borovi{\v c}ka} J.,  {Spurn{\'y}} P.,    {Koten} P.,  2007, \aap, 473, 661

\bibitem[\protect\citeauthoryear{{Brown}, {Jones}, {Weryk} \&
  {Campbell-Brown}}{{Brown} et~al.}{2004}]{bro04}
{Brown} P.,  {Jones} J.,  {Weryk} R.~J.,    {Campbell-Brown} M.~D.,  2004,
  Earth Moon and Planets, 95, 617

\bibitem[\protect\citeauthoryear{{Brown}, {Weryk}, {Wong} \& {Jones}}{{Brown}
  et~al.}{2008}]{bro08}
{Brown} P.,  {Weryk} R.~J.,  {Wong} D.~K.,    {Jones} J.,  2008, \icarus, 195,
  317

\bibitem[\protect\citeauthoryear{{Brown}, {Wong}, {Weryk} \& {Wiegert}}{{Brown}
  et~al.}{2010}]{bro10}
{Brown} P.,  {Wong} D.~K.,  {Weryk} R.~J.,    {Wiegert} P.,  2010, \icarus,
  207, 66

\bibitem[\protect\citeauthoryear{{Campbell-Brown}, {Brown}, {Wiegert}, {Arlt}
  \& {Vaubaillon}}{{Campbell-Brown} et~al.}{2005}]{cam05}
{Campbell-Brown} M.,  {Brown} P.,  {Wiegert} P.,  {Arlt} R.,    {Vaubaillon}
  J.,  2005, Central Bureau Electronic Telegrams, 255, 1

\bibitem[\protect\citeauthoryear{{Campbell-Brown}, {Vaubaillon}, {Brown},
  {Weryk} \& {Arlt}}{{Campbell-Brown} et~al.}{2006}]{cam06}
{Campbell-Brown} M.,  {Vaubaillon} J.,  {Brown} P.,  {Weryk} R.~J.,    {Arlt}
  R.,  2006, \aap, 451, 339

\bibitem[\protect\citeauthoryear{{Campbell-Brown} \&
  {Koschny}}{{Campbell-Brown} \& {Koschny}}{2004}]{cam04}
{Campbell-Brown} M.~D.,  {Koschny} D.,  2004, \aap, 418, 751

\bibitem[\protect\citeauthoryear{{Ceplecha}, {Borovi{\v c}ka}, {Elford},
  {Revelle}, {Hawkes}, {Porub{\v c}an} \& {{\v S}imek}}{{Ceplecha}
  et~al.}{1998}]{cep98}
{Ceplecha} Z.,  {Borovi{\v c}ka} J.,  {Elford} W.~G.,  {Revelle} D.~O.,
  {Hawkes} R.~L.,  {Porub{\v c}an} V.,    {{\v S}imek} M.,  1998, \ssr, 84, 327

\bibitem[\protect\citeauthoryear{{Cervera}, {Elford} \& {Steel}}{{Cervera}
  et~al.}{1997}]{cer97}
{Cervera} M.~A.,  {Elford} W.~G.,    {Steel} D.~I.,  1997, Radio Science, 32,
  805

\bibitem[\protect\citeauthoryear{{Davidson}}{{Davidson}}{1915}]{dav15}
{Davidson} R.~M.,  1915, JBAA, 25, 292

\bibitem[\protect\citeauthoryear{{Davies} \& {Lovell}}{{Davies} \&
  {Lovell}}{1955}]{dav55}
{Davies} J.~G.,  {Lovell} A.~C.~B.,  1955, \mnras, 115, 23

\bibitem[\protect\citeauthoryear{{Denning}}{{Denning}}{1926}]{den27}
{Denning} W.~F.,  1926, \mnras, 87, 104

\bibitem[\protect\citeauthoryear{{Emel'Ianenko}}{{Emel'Ianenko}}{1992}]{eme92}
{Emel'Ianenko} V.~V.,  1992, Celestial Mechanics and Dynamical Astronomy, 54,
  91

\bibitem[\protect\citeauthoryear{{Fiocco} \& {Colombo}}{{Fiocco} \&
  {Colombo}}{1964}]{fio64}
{Fiocco} G.,  {Colombo} G.,  1964, \jgr, 69, 1795

\bibitem[\protect\citeauthoryear{{Fisher}}{{Fisher}}{1934}]{fis34}
{Fisher} W.~J.,  1934, Harvard College Observatory Bulletin, 894, 15

\bibitem[\protect\citeauthoryear{{Hey}, {Parsons} \& {Stewart}}{{Hey}
  et~al.}{1947}]{hey47}
{Hey} J.~S.,  {Parsons} S.~J.,    {Stewart} G.~S.,  1947, \mnras, 107, 176

\bibitem[\protect\citeauthoryear{{Hocking}, {Fuller} \& {Vandepeer}}{{Hocking}
  et~al.}{2001}]{hoc01}
{Hocking} W.~K.,  {Fuller} B.,    {Vandepeer} B.,  2001, Journal of Atmospheric
  and Solar-Terrestrial Physics, 63, 155

\bibitem[\protect\citeauthoryear{{Hughes} \& {Thompson}}{{Hughes} \&
  {Thompson}}{1973}]{hug73}
{Hughes} D.~W.,  {Thompson} D.~A.,  1973, \mnras, 163, 3P

\bibitem[\protect\citeauthoryear{{Imoto} \& {Hasegawa}}{{Imoto} \&
  {Hasegawa}}{1958}]{imo58}
{Imoto} S.,  {Hasegawa} I.,  1958, Smithsonian Contributions to Astrophysics,
  2, 131

\bibitem[\protect\citeauthoryear{{Jacchia}, {Kopal} \& {Millman}}{{Jacchia}
  et~al.}{1950}]{jac50}
{Jacchia} L.~G.,  {Kopal} Z.,    {Millman} P.~M.,  1950, \apj, 111, 104

\bibitem[\protect\citeauthoryear{{Jenniskens}}{{Jenniskens}}{1995}]{jen95}
{Jenniskens} P.,  1995, \aap, 295, 206

\bibitem[\protect\citeauthoryear{{Jenniskens}}{{Jenniskens}}{2006}]{jen06}
{Jenniskens} P.,  2006, {Meteor Showers and their Parent Comets}

\bibitem[\protect\citeauthoryear{{Jenniskens}, {Barentsen} \&
  {Yrjola}}{{Jenniskens} et~al.}{2011}]{jen11}
{Jenniskens} P.,  {Barentsen} G.,    {Yrjola} I.,  2011, Central Bureau
  Electronic Telegrams, 2862, 1

\bibitem[\protect\citeauthoryear{{Jones}, {Brown}, {Ellis}, {Webster},
  {Campbell-Brown}, {Krzemenski} \& {Weryk}}{{Jones} et~al.}{2005}]{jon05}
{Jones} J.,  {Brown} P.,  {Ellis} K.~J.,  {Webster} A.~R.,  {Campbell-Brown}
  M.,  {Krzemenski} Z.,    {Weryk} R.~J.,  2005, \planss, 53, 413

\bibitem[\protect\citeauthoryear{{Kero}, {Fujiwara}, {Abo}, {Szasz} \&
  {Nakamura}}{{Kero} et~al.}{2012}]{ker12}
{Kero} J.,  {Fujiwara} Y.,  {Abo} M.,  {Szasz} C.,    {Nakamura} T.,  2012,
  \mnras, 424, 1799

\bibitem[\protect\citeauthoryear{{Kronk}}{{Kronk}}{2008}]{kro08}
{Kronk} G.~W.,  2008, {Cometography}

\bibitem[\protect\citeauthoryear{{Langbroek}}{{Langbroek}}{1997}]{lan97}
{Langbroek} M.,  1997, WGN, Journal of the International Meteor Organization,
  25, 37

\bibitem[\protect\citeauthoryear{{Lindblad}}{{Lindblad}}{1987}]{lin87}
{Lindblad} B.~A.,  1987, \aap, 187, 928

\bibitem[\protect\citeauthoryear{{Lovell}, {Banwell} \& {Clegg}}{{Lovell}
  et~al.}{1947}]{lov47}
{Lovell} A.~C.~B.,  {Banwell} C.~J.,    {Clegg} J.~A.,  1947, \mnras, 107, 164

\bibitem[\protect\citeauthoryear{{Maley} \& {Saulietis}}{{Maley} \&
  {Saulietis}}{1972}]{mal72}
{Maley} P.~D.,  {Saulietis} I.,  1972, \iaucirc, 2452, 1

\bibitem[\protect\citeauthoryear{{Maslov}}{{Maslov}}{2011}]{mas11}
{Maslov} M.,  2011, WGN, Journal of the International Meteor Organization, 39,
  64

\bibitem[\protect\citeauthoryear{{Massey}, {Burhop} \& {Gilbody}}{{Massey}
  et~al.}{1971}]{mas71}
{Massey} H.~S.~W.,  {Burhop} E.~H.~S.,    {Gilbody} H.~B.,  1971, {Electronic
  and ionic impact phenomena. Vol.\_3: Slow collisions of heavy particles.}

\bibitem[\protect\citeauthoryear{{McIntosh}}{{McIntosh}}{1968}]{mci68}
{McIntosh} B.~A.,  1968, in {Kresak} L.,  {Millman} P.~M.,  eds, Physics and
  Dynamics of Meteors Vol.~33 of IAU Symposium, {Meteor Mass Distribution from
  Radar Observations}.
p.~343

\bibitem[\protect\citeauthoryear{{McKinley}}{{McKinley}}{1961}]{mck61}
{McKinley} D.~W.~R.,  1961, {Meteor science and engineering.}

\bibitem[\protect\citeauthoryear{{Nijland} \& {van der Bilt}}{{Nijland} \& {van
  der Bilt}}{1935}]{nij35}
{Nijland} A.~A.,  {van der Bilt} J.,  1935, \bain, 7, 248

\bibitem[\protect\citeauthoryear{{Sato}}{{Sato}}{2003}]{sat03}
{Sato} M.,  2003, WGN, Journal of the International Meteor Organization, 31, 59

\bibitem[\protect\citeauthoryear{{Simek}}{{Simek}}{1994}]{sim94}
{Simek} M.,  1994, \aap, 284, 276

\bibitem[\protect\citeauthoryear{{Simek} \& {McIntosh}}{{Simek} \&
  {McIntosh}}{1968}]{sim68}
{Simek} M.,  {McIntosh} B.~A.,  1968, in {Kresak} L.,  {Millman} P.~M.,  eds,
  Physics and Dynamics of Meteors Vol.~33 of IAU Symposium, {Meteor Mass
  Distribution from Underdense-Trail Echoes}.
p.~362

\bibitem[\protect\citeauthoryear{{Spalding}}{{Spalding}}{1985}]{spa85}
{Spalding} G.~H.,  1985, Journal of the British Astronomical Association, 95,
  211

\bibitem[\protect\citeauthoryear{{Stewart}, {Ference}, {Slattery} \&
  {Zahl}}{{Stewart} et~al.}{1947}]{ste47}
{Stewart} J.~Q.,  {Ference} M.,  {Slattery} J.~J.,    {Zahl} H.~A.,  1947, \aj,
  52, 158

\bibitem[\protect\citeauthoryear{{Vaubaillon}, {Colas} \& {Jorda}}{{Vaubaillon}
  et~al.}{2005}]{vau05}
{Vaubaillon} J.,  {Colas} F.,    {Jorda} L.,  2005, \aap, 439, 751

\bibitem[\protect\citeauthoryear{{Vaubaillon}, {Koten}, {Gerding}, {Johannink},
  {Langbroek}, {Latteck}, {Brown} \& {Jenniskens}}{{Vaubaillon}
  et~al.}{2011}]{vau11b}
{Vaubaillon} J.,  {Koten} P.,  {Gerding} M.,  {Johannink} C.,  {Langbroek} M.,
  {Latteck} R.,  {Brown} P.,    {Jenniskens} P.,  2011, Central Bureau
  Electronic Telegrams, 2862, 2

\bibitem[\protect\citeauthoryear{{Vaubaillon}, {Watanabe}, {Sato}, {Horii} \&
  {Koten}}{{Vaubaillon} et~al.}{2011}]{vau11}
{Vaubaillon} J.,  {Watanabe} J.,  {Sato} M.,  {Horii} S.,    {Koten} P.,  2011,
  WGN, Journal of the International Meteor Organization, 39, 59

\bibitem[\protect\citeauthoryear{{Verniani}}{{Verniani}}{1973}]{ver73}
{Verniani} F.,  1973, \jgr, 78, 8429

\bibitem[\protect\citeauthoryear{{Watson}}{{Watson}}{1939}]{wat39}
{Watson} F.,  1939, \nat, 144, 482

\bibitem[\protect\citeauthoryear{{Watson} Jr.}{{Watson}}{1934}]{wat34}
{Watson} Jr. F.,  1934, Harvard College Observatory Bulletin, 895, 9

\bibitem[\protect\citeauthoryear{{Watson}}{{Watson}}{1946}]{wat46}
{Watson} F.~G.,  1946, \skytel, 5, 3

\bibitem[\protect\citeauthoryear{{Weiss}}{{Weiss}}{1961}]{wei61}
{Weiss} A.~A.,  1961, Australian Journal of Physics, 14, 102

\bibitem[\protect\citeauthoryear{{Weryk} \& {Brown}}{{Weryk} \&
  {Brown}}{2012}]{wer12}
{Weryk} R.~J.,  {Brown} P.~G.,  2012, \planss, 62, 132

\end{thebibliography}

\label{lastpage}

\end{document}